\documentclass[runningheads]{llncs}

\usepackage[T1]{fontenc}
\usepackage{amsmath, amsfonts, amssymb}
\usepackage{tikz} \usetikzlibrary{calc, positioning, arrows.meta}
\usepackage{graphicx} \graphicspath{ {./figures/} }
\usepackage{hyperref}
\usepackage{xcolor}  
\usepackage{cleveref}
\usepackage{nicefrac}
\usepackage{algorithm}
\usepackage{algorithmic}
\usepackage{bm}
\usepackage{microtype}
\usepackage{orcidlink} \renewcommand{\orcidID}[1]{\orcidlink{#1}}

\newcommand{\R}{\mathbb{R}}
\newcommand{\N}{\mathbb{N}}
\newcommand{\Z}{\mathbb{Z}}
\newcommand{\psep}[1]{P_{\text{SEP}}^{#1}}
\newcommand{\bb}{{\normalfont\textsc{bb}}}
\newcommand{\tsp}{\text{TSP}}
\newcommand{\sep}{\text{SEP}}
\newcommand{\opt}{\text{OPT}}
\newcommand{\gap}{\text{Gap}}
\newcommand{\gapp}{\ifmmode\text{Gap}^+\else\text{Gap}\textsuperscript{$+$}\fi}
\newcommand{\optp}{\ifmmode\text{OPT}^+\else\text{OPT}\textsuperscript{$+$}\fi}
\newcommand{\opti}{\ifmmode\text{OPT}^\text{I}\else\text{OPT}\textsuperscript{I}\fi}
\newcommand{\optii}{\ifmmode\text{OPT}^\text{II}\else\text{OPT}\textsuperscript{II}\fi}
\newcommand{\dual}{\mathcal{D}\,}
\newcommand{\cS}{\mathcal{S}}
\newcommand{\cF}{\mathcal{F}}
\newcommand{\cA}{\mathcal{A}}
\newcommand{\cW}{\mathcal{W}}
\newcommand{\gb}{\text{GB}}

\newcommand{\ee}{{\normalfont\textsc{e}}}

\renewcommand{\setminus}{\!\smallsetminus\!}

\title{The Integrality Gap of the Traveling Salesman Problem is \texorpdfstring{$\nicefrac{4}{3}$}{4/3} if the LP Solution Has at Most \texorpdfstring{$n + 6$}{n + 6} Non-Zero Components}
\titlerunning{The Integrality Gap of the Traveling Salesman Problem}

\author{Tullio Villa\inst{1,3}\orcidID{0009-0003-4669-4781}
    \and Eleonora Vercesi\inst{2,3}\orcidID{0000-0002-1621-2484}
    \and Janos Barta\inst{1,3}\orcidID{0000-0003-1577-5637}
    \and Monaldo Mastrolilli\inst{1,3}\orcidID{0000-0002-2948-9749}}
\authorrunning{T. Villa et al.}

\institute{
    Scuola Universitaria Professionale della Svizzera Italiana (SUPSI)
    \and
    Università della Svizzera Italiana (USI)
    \and
    Dalle Molle Institute for Artificial Intelligence USI-SUPSI
}

\begin{document}

%%%%%%%%%%%%%%%%%%%%%%%%%
% Title page
%%%%%%%%%%%%%%%%%%%%%%%%%
\maketitle

\begin{abstract}
We address the classical Dantzig–Fulkerson–Johnson formulation of the symmetric metric Traveling Salesman Problem and study the integrality gap of its linear relaxation, namely the Subtour Elimination Problem (\sep).
This integrality gap is conjectured to be $\nicefrac{4}{3}$.
We prove that, when solving a problem on $n$ nodes, if the optimal \sep\ solution has at most $n + 6$ non-zero components, then the conjecture is true.
To establish this result, we devise a new methodology that combines theoretical analysis and computational verification.

\bigskip

\textbf{Publication note.} This paper appeared in the Proceedings of IPCO 2026.
For citation purposes, please cite the published conference version~\cite{art:VilVerBarMas:IGTSP6Edges}.

% \keywords{Symmetric Traveling Salesman Problem \and Linear Programming Relaxation \and Integrality Gap}
\end{abstract}

%%%%%%%%%%%%%%%%%%%%%%%%%
% Introduction
%%%%%%%%%%%%%%%%%%%%%%%%%
\section{Introduction} \label{sec:introduction}
\setcounter{page}{1} 

Let $K_n = (V_n, E_n)$ be the complete graph on $n$ nodes, and let $\bm{c} \in \R^{E_n}$ be a non-negative cost vector.
The \emph{Traveling Salesman Problem} (\tsp) consists of finding a Hamiltonian tour of minimum total cost.
This paper focuses exclusively on the \emph{symmetric} formulation, where the graph $K_n$ is undirected, hence we use $ij$ and $ji$ interchangeably to denote the edge connecting nodes $i$ and $j$.
Additionally, we only consider \emph{metric} cost vectors, satisfying the triangle inequalities.
Henceforth, we simply use the term \tsp\ to indicate the symmetric metric \tsp.

The \tsp\ is typically formulated as an Integer Linear Program (ILP).
Among the most prominent formulations is the one of Dantzig-Fulkerson-Johnson (DFJ) \cite{dantzig1954solution}, whose associated relaxed Linear Problem (LP) is known in the literature as the Subtour Elimination Problem (\sep), or Held-Karp relaxation:
$$
\text{DFJ}: \
\min_{\bm{x} \in \psep{n} \cap \Z^{E_n}} \sum_{e \in E_n} c_e x_e \ ,
\hspace{1cm}
\sep: \
\min_{\bm{x} \in \psep{n}} \sum_{e \in E_n} c_e x_e \ .
$$
The polytope on which they are described is the \emph{subtour elimination polytope}:
{\small
$$
\psep{n} := \bigg\{ \bm{x} \! \in \! \R^{E_n} \mid 
\sum_{e \in \delta(v)} \!\!\! x_e = 2 \ \ \forall v \! \in \! V_n, \
\sum_{e \in \delta(S)} \!\!\! x_e \geq 2 \ \ \forall S \! \in \! \cS, \
 0 \leq x_e \leq 1 \ \ \forall e \! \in \! E_n
\bigg\} \ ,
$$}%
where $\cS := \{ S \subseteq V_n \mid 3 \leq |S| \leq n-3 \}$ and,for any subset $S \subseteq V_n$, $\delta(S)$ is the set of edges having exactly one node in $S$
\footnote{The brackets denoting singletons are omitted for brevity, e.g., $v = \{ v \}$; this abuse of notation is used throughout this work, when the context does not lead to ambiguity.}.
The term \emph{subtour elimination constraints} refers to the second set of constraints in the definition of $\psep{n}$.

For a given cost vector $\bm{c}$, we denote the optimal solutions of the integral and relaxed problem by $\tsp(\bm{c})$ and $\sep(\bm{c})$, respectively.
The \emph{integrality gap} is the ratio of these two values; it can be evaluated for a single instance cost $\bm{c}$, as $\alpha(\bm{c})$, or considered in a worst case scenario, as $\alpha$:
$$
\alpha(\bm{c}) :=  \frac{\tsp(\bm{c})}{\sep(\bm{c})} \ , \qquad
\alpha := \sup_{\bm{c} \text{ metric}} \frac{\tsp(\bm{c})}{\sep(\bm{c})} \ .
$$
An alternative granularity of the integrality gap can also be defined by focusing on a specific vertex $\bm{x} \in \psep{n}$.
We denote this by $\gap(\bm{x})$, and it is given by:
$$
\gap(\bm{x}) := \sup \left\{
\frac{\tsp(\bm{c})}{\sep(\bm{c})} \ \mid \ \bm{c} \text{ metric }, \ \bm{x} \in \arg\min \sep(\bm{c})
\right\} \ .
$$

The exact value of $\alpha$ is currently unknown; in order to determine it, various approaches have been explored, with increasing interest over the past decade. 
Wolsey \cite{art:Wol:HeusLPB&B} proposed an upper bound of $\nicefrac{3}{2}$, which remains the best known to date. 
Already in 1995, Goemans \cite{goemans1995worst} conjectured that the integrality gap was equal to $\nicefrac{4}{3}$, and despite several attempts, nobody was able to disprove that conjecture. 
Later, \cite{art:BenBoy:IGSmallTSP,art:BoyEll:ExtrSEP} exploited exhaustive enumeration of the vertices of $\psep{n}$ to compute the exact value of the integrality gap for \tsp\ instances with $n \leq 12$. 
Other than that, only results for specific subclasses of instances are available in the literature.
Schalekamp,  Williamson, and van Zuylen \cite{schalekamp20142} conjectured that the maximum integrality gap is attained on \emph{half-integer} instances, that is, cost vectors whose optimal \sep\ solutions have all the entries in $\{0, \nicefrac{1}{2}, 1\}$.
In recent years, promising lines of research have examined subclasses of \sep\ solutions $\bm{x}$ characterized by specific properties of the so called \emph{support graph}, defined as the undirected weighted graph $G_{\bm{x}} = (V_n, E_{\bm{x}})$ such that $ij \in E_{\bm{x}}$ $\Leftrightarrow$ $x_{ij} > 0$, \ and the weight on edge $ij$ is given by $x_{ij}$.
Notice that the number of edges in the support graph $G_{\bm{x}}$ is, by definition, the number of non-zero components of $\bm{x}$.
Boyd and Carr \cite{art:BoyCar:TSP2Match} proved that the integrality gap is $\nicefrac{4}{3}$ when the support graph of the \sep\ solution contains disjoint $\nicefrac{1}{2}$ triangles. 
For graph-\tsp\ instances%
\footnote{Graph-\tsp\ is a subclass of metric \tsp\ where the distance between the nodes is computed as the minimum number of edges separating them.},
Boyd et al. \cite{art:BoySitSteSto:TSPCubic} proved that the $\nicefrac{4}{3}$-conjecture holds for cubic graphs.
Mömke and Svensson \cite{momke_removing_2016} improved this result showing that the integrality gap is $\nicefrac{4}{3}$ for graph-\tsp\
restricted either to half-integral solutions or to a class of graphs that contains subcubic and claw-free graphs.
Boyd and Sebő \cite{boyd2021salesman} proved that the integrality gap is at most $\sim 1.4286$ for instances having as \sep\ solution one of the so-called Boyd-Carr points (firstly defined in \cite{art:BoyCar:TSP2Match}).
Jin et al. \cite{art:JinKleWil:CycleCutTSP} proved the $\nicefrac{4}{3}$-conjecture for instances having as \sep\ solution cycle-cut points, that is, points for which every non-singleton tight set can be written as the union of two tight sets.
With a breakthrough result \cite{art:KarKleGha:slightlyApproxTSP}, Karlin, Klein, and Oveis Gharan gave an approximation algorithm for the general \tsp; building on this work, the same authors were able to prove an integrality gap smaller than $\nicefrac{3}{2}$ in the half-integer case \cite{art:KarKleGha:halfTSP}.
This factor was improved in \cite{gupta2024matroid} and subsequently in the as-yet unpublished work \cite{klein2025dual}.

Different approaches have also been attempted to improve the lower bound. 
In \cite{art:BenBoy:IGSmallTSP,hougardy2014integrality,hougardy2021hard,zhong2025lower}, families of \tsp\ instances with high integrality gap, asymptotically tending to $\nicefrac{4}{3}$ were shown.
In \cite{vercesi2023generation}, the authors proposed an approach to \emph{heuristically} generate instances with a high integrality gap: no instance with an integrality gap greater than $\nicefrac{4}{3}$ was found.

%%%%%%%%%% Contribution ad outlook
\paragraph{Contribution and outlook.}
This work presents both a theoretical contribution and a novel methodology that opens new research directions in the study of the integrality gap.
We prove the $\nicefrac{4}{3}$-conjecture for all the vertices of $\psep{n}$ whose support graph contains at most $n + 6$ edges.
Remarkably, this class is rich enough to contain vertices that are not covered by any previously known result in the literature.
\Cref{fig:example_vertex} shows an example of such a vertex.

\begin{figure}
    \centering
    \begin{tikzpicture}[x=6mm, y=6mm]
    \node at (0, 0) (a) {};
    \node at (4, 0) (b) {};
    \node at (4, 4) (c) {};
    \node at (0, 4) (d) {};
    \node at (1, 1) (e) {};
    \node at (3, 1) (f) {};
    \node at (3, 3) (g) {};
    \node at (1, 3) (h) {};
          
    \foreach \vname in {a, b, c, d, e, f, g, h}{
        \fill (\vname) circle (2pt);
    }
    
    \draw plot coordinates {(a) (h) (d) (a)}
          plot coordinates {(b) (e) (f) (b)}
          plot coordinates {(b) (c) (g) (h)};
    
    \foreach \p/\q in {a/b, c/d, e/h, f/g}{
        \fill ($(\p)!1/3!(\q)$) circle (2pt);
        \fill ($(\p)!2/3!(\q)$) circle (2pt);
        \draw plot coordinates {(\p) ($(\p)!1/3!(\q)$)}
              plot coordinates {($(\p)!2/3!(\q)$) (\q)};
        \draw[dashed] plot coordinates {($(\p)!1/3!(\q)$) ($(\p)!2/3!(\q)$)};
    }
    
    \node[below=-1pt] at ($(a)!1/6!(b)$) {\scriptsize $1$};
    \node[below=-1pt] at ($(a)!5/6!(b)$) {\scriptsize $1$};
    \node[above=-1pt] at ($(c)!1/6!(d)$) {\scriptsize $1$};
    \node[above=-1pt] at ($(c)!5/6!(d)$) {\scriptsize $1$};
    \node[right=-1pt] at ($(e)!1/6!(h)$) {\scriptsize $1$};
    \node[right=-1pt] at ($(e)!5/6!(h)$) {\scriptsize $1$};
    \node[left=-1pt]  at ($(f)!1/6!(g)$) {\scriptsize $1$};
    \node[left=-1pt]  at ($(f)!5/6!(g)$) {\scriptsize $1$};

    \node[left=-1pt]  at (0, 2) {\scriptsize $\nicefrac{2}{3}$};
    \node[right=-1pt] at (4, 2) {\scriptsize $\nicefrac{1}{3}$};
    \node[below=-1pt] at (2, 3) {\scriptsize $\nicefrac{1}{3}$};
    \node[above=-1pt] at (2, 1) {\scriptsize $\nicefrac{2}{3}$};
    
    \node[right=-1pt] at (0.2, 0.6) {\scriptsize $\nicefrac{1}{3}$};
    \node[right=-1pt] at (0.5, 3.5) {\scriptsize $\nicefrac{1}{3}$};
    \node[left=-1pt]  at (3.5, 3.5) {\scriptsize $\nicefrac{2}{3}$};
    \node[above=-1pt] at (3.5, 0.5) {\scriptsize $\nicefrac{1}{3}$};
    \node[left=-1pt]  at (2, 0.5)   {\scriptsize $\nicefrac{1}{3}$};
\end{tikzpicture}
    \caption{Example of a vertex considered in this work but not in previous literature.
    Dashed edges may be replaced by arbitrarily long paths of edges of weight $1$.}
    \label{fig:example_vertex}
\end{figure}

To establish our result, we define, for each integer $k$, the family $\cF_k$ which comprises, across all polytopes $\psep{n}$ for $n \in \N$, the vertices whose support graphs contain exactly $n + k$ edges.
Although the family $\cF_k$ is infinite, we identify a finite subset $\cA_k$ of its elements, which we call \emph{ancestors}, that allows us to make overall considerations on the entire family.
We devise the \emph{\gap-Bounding (\gb) algorithm} and prove that its application on these ancestors returns upper bounds on the integrality gap of all the costs whose \sep\ solution is a vertex of $\cF_k$.
The application of the \gb\ algorithm on the elements of $\cA_k$ with $k \leq 6$ yields a computer-aided proof of the desired result.

The methodology presented constitutes in itself a novel way to approach the integrality gap problem, combining theoretical analysis and computational verification.
Unlike earlier computational proofs (see \cite{art:BenBoy:IGSmallTSP,art:BoyEll:ExtrSEP}) that were limited to enumerating vertices for a fixed dimension $n$, our framework fixes a small parameter $k$ and derives results valid for \emph{all} $n$, hereby achieving a form of universality through a computational approach.
To the best of our knowledge, this is the first time that a computer-aided proof is implied in bounding the integrality gap for \emph{infinitely many} vertices.

We conclude by noting that the vast majority of instances used in the literature to establish lower bounds on the integrality gap are special cases of our result (see, e.g., \cite{art:BenBoy:IGSmallTSP,hougardy2014integrality,hougardy2021hard,zhong2025lower}%
\footnote{Counting nodes and edges in the constructions, one may see that the instances in~\cite{art:BenBoy:IGSmallTSP,hougardy2014integrality,zhong2025lower} belong to $\cF_3$ and the instances in~\cite{hougardy2021hard} belong to $\cF_6$.}).
Other families, however, fall outside the scope of our analysis: the donut instances proposed in \cite{boyd2021salesman} have a non-constant surplus of edges over nodes, and the lower bounds on their integrality gap provided in the article converge to $\nicefrac{4}{3}$.
Interestingly, for each donut instance, the mentioned lower bound is consistently dominated by the integrality gap of an instance of $\cF_3$ with the same number of nodes%
\footnote{It can be checked comparing the explicit formulas for the lower bounds on the integrality gap in \cite{art:BenBoy:IGSmallTSP} and \cite{boyd2021salesman}.}.
This observation aligns with the empirical evidence that, for a small number of nodes $n \leq 12$, the value of the integrality gap of a vertex correlates with the number of edges in its support graph (more details in \Cref{sec:conclusion}).
Taken together, these insights motivate the following hypothesis, which highlights the significance of the class of vertices considered in this work.

\begin{center}
    \begin{minipage}{0.95\textwidth}
        \textbf{Few Edges Hypothesis}.
        \textit{For every $n$, among the vertices of $\psep{n}$, the maximum integrality gap is realized on a vertex with the minimum possible number of edges in the support graph, that is $n + 3$.}
    \end{minipage}
\end{center}

In summary, our work contributes to three different fronts.
\begin{enumerate}
    \item It establishes the $\nicefrac{4}{3}$ conjecture for newly identified infinite classes of vertices.
    \item It demonstrates the potential of computer-aided proofs to advance our understanding of the integrality gap across infinite vertex classes, proposing a flexible framework that can be adapted to other contexts.
    \item It opens the compelling research direction of investigating the correlation between the integrality gap and the number of edges in the support graph.
    A proof of the Few Edges Hypothesis, together with the contribution of this work, would result in definitively solving the $\nicefrac{4}{3}$ conjecture.
\end{enumerate}

%%%%%%%%%% Outline
\paragraph{Outline.}
\Cref{sec:background} introduces key definitions from the literature.
\Cref{sec:families} presents the concept of ancestors together with the procedure to retrieve them.
\Cref{sec:gapp} provides the theoretical basis for the \gb\ algorithm, whose design is detailed in \Cref{sec:gap_bounding}.
\Cref{sec:comp_results} explains how to apply the \gb\ algorithm to bound the integrality gap for vertices with up to $n + 6$ non-zero components.
\Cref{sec:conclusion} explores future research directions.
Proofs and technical details are included in the appendix.

%%%%%%%%%%%%%%%%%%%%%%%%%
% Background
%%%%%%%%%%%%%%%%%%%%%%%%%
\section{Background material} \label{sec:background}

%%%%%%%%%% Preliminaries from graph theory
\subsection{Preliminaries from graph theory}

Before giving the list of definitions used in this work, we clarify that, to avoid confusion, given a graph $G = (V, E)$, the term \emph{node} refers to the elements of $V$; the term \emph{vertex} is reserved exclusively for the extreme points of $\psep{n}$.

\begin{definition}[Hamiltonian walk, Hamiltonian tour] \label{def:walk}
    Let $G$ be an undirected graph.
    A \emph{Hamiltonian walk} is a closed walk in $G$ that visits every node at least once, possibly traversing the same edge multiple times.
    If every node is visited exactly once, we call the walk \emph{Hamiltonian tour}.
    For a given walk $\bm{w}$, $w_{ij} \in \N$ denotes the \emph{multiplicity} of $ij$ in $\bm{w}$, that is, the number of times the walk $\bm{w}$ uses the edge $ij$
    (the analogous notation $t_{ij} \in \{ 0, 1 \}$ is used for a tour $\bm{t}$).
\end{definition}

For brevity, in what follows the terms \emph{walk} and \emph{tour} always indicate Hamiltonian walks and tours.
Moreover, we use the same literal to denote both the walk $\bm{w}$ (respectively the tour $\bm{t}$) and its \emph{characteristic vector}, that is, the vector of $w_{ij}$ (respectively $t_{ij}$) entries.
When the underlying graph $G$ is not specified, we assume it is the complete one.

\begin{remark} \label{rem:walk_mult}
    Since we have interest in the shortest possible walks $\bm{w}$, we assume that every edge $ij$ is traversed at most twice (i.e. $w_{ij} \in \{ 0, 1, 2 \}$)%
    \footnote{It is well known (see, e.g. \cite{art:CorFonNad:TSPGraph}) that, when an edge is used strictly more than twice by a walk, it is possible to get rid of two copies of that edge to obtain a shorter walk.}.
    Notice that, under this assumption, the number of walks considered on a graph becomes finite.
\end{remark}

%%%%%%%%%% Preliminaries from the literature on the integrality gap
\subsection{Preliminaries from the literature on the integrality gap}

In this section, we collect from the literature (mainly from \cite{art:BenBoy:IGSmallTSP}) the definitions and results are used in this manuscript.

\begin{theorem}[\cite{art:BenBoy:IGSmallTSP,art:BoyPul:SEPPoly}] \label{thm:max_edges}
    Let $\bm{x} \in \psep{n}$ be a vertex.
    Then \ $|E_{\bm{x}}| \leq 2n - 3$.
\end{theorem}

\begin{definition}[$1$-edge, $1$-path, from~\cite{art:BenBoy:IGSmallTSP}]
    Let $\bm{x} \in \psep{n}$ and let $e$ be an edge of $K_n$; $e$ is called \emph{$1$-edge} of $\bm{x}$ if \ $x_e = 1$.
    When $\bm{x}$ is fractional (that is, not a tour), we call \emph{$1$-path} of $\bm{x}$ a maximal path of $1$-edges in the support graph $G_{\bm{x}}$; the nodes of degree $2$ in the $1$-path are called \emph{internal nodes}, the two remaining nodes are called \emph{end nodes}.
\end{definition}

\begin{theorem}[\cite{art:BenBoy:IGSmallTSP}] \label{thm:3_1paths}
    Let $\bm{x} \in \psep{n}$ be a fractional vertex.
    Then $\bm{x}$ has at least three distinct $1$-paths.
\end{theorem}

We now introduce a construction presented in~\cite{art:BenBoy:IGSmallTSP} and give it a name using the authors' initials.

\begin{definition}[\bb-move, from \cite{art:BenBoy:IGSmallTSP}] \label{def:BBmove}
    Let $\bm{x} \in \psep{n}$ and let $ab$ be one of its $1$-edges.
    We call \emph{\bb-move} the construction of a new point $\bm{x}' \in \psep{n+1}$ defined as follows, where $w$ is the new node added ($V_{n+1} = V_n \cup \{ w \}$):
    $$
    \setlength{\arraycolsep}{15pt}
    \begin{array}{ll}
        x'_{ab} = 0 \ ,
            & x'_e = 0 \quad \forall e \in \delta(w) \setminus \{ aw, wb \} \ , \\
        x'_{aw} = x'_{wb} = 1 \ ,
            & x'_e = x_e \quad \forall e \notin \delta(w) \cup \{ ab \} \ .
    \end{array}
    $$
    We denote this construction by \ $\bb(\bm{x}, ab) := \bm{x}'$. \
    When it is clear from the context or not strictly necessary, we omit the edge $ab$ in the notation: \ $\bb(\bm{x})$.
\end{definition}

\begin{theorem}[\cite{art:BenBoy:IGSmallTSP}] \label{thm:bb_move}
    Let $\bm{x} \in \psep{n}$ and let $ab$ be one of its $1$-edges.
    Then $\bb(\bm{x}, ab)$ is a vertex of $\psep{n+1}$ if and only if $\bm{x}$ is a vertex of $\psep{n}$.
\end{theorem}

%%%%%%%%%%%%%%%%%%%%%%%%%
% Families of vertices with a given number of edges
%%%%%%%%%%%%%%%%%%%%%%%%%
\section{Families of vertices with a given number of edges} \label{sec:families}

In this section, we study the vertices of $\psep{n}$ considering, in their support graph, the relation between the number of edges and the number of nodes.

\begin{lemma} \label{lem:min_edges}
    Let $\bm{x}$ be a fractional vertex of $\psep{n}$.
    Then \ $|E_{\bm{x}}| \geq n + 3$. 
\end{lemma}

Our aim now is to describe, for a given $k$, all possible fractional vertices with $n$ nodes and $n + k$ edges.
For this purpose, we define the families of vertices $\cF_k$:
$$
\cF_k := \{ \bm{x} \text{ fractional vertex} \ \mid \
\bm{x} \in \psep{n} \text{ for some } n, \
|E_{\bm{x}}| = n + k \} \ .
$$
Notice that \Cref{lem:min_edges} implies $k \geq 3$, hence $\cF_1 = \cF_2 = \emptyset$.
Observe also that a \bb-move increases both the number of nodes and edges by $1$, thus preserving their difference $k$.
Therefore, in virtue of \Cref{thm:bb_move}, a vertex is in $\cF_k$ if and only if its image under a \bb-move is in $\cF_k$ as well. 
This brings us to define the \emph{ancestors}, the vertices of $\cF_k$ that can not be obtained as results of a \bb-move, that is, vertices without internal nodes in the $1$-paths.

\begin{definition}[Ancestor of order $k$]
    An \emph{ancestor of order $k$} is a vertex $\bm{x}$ of $\cF_k$ with no node of degree $2$.
    We denote the set of ancestors of order $k$ as
    $$
    \cA_k := \{ \bm{x} \in \cF_k \mid \bm{x} \text{ has no node of degree } 2 \} \ .
    $$
\end{definition}

\begin{definition}[Successor]
    Given a vertex $\bm{x}$, we call \emph{successor} of $\bm{x}$ any vertex $\bm{x}'$ obtained by sequential applications of the \bb-move.
\end{definition}

We can thus recover the whole family $\cF_k$ by taking the elements of $\cA_k$ and replacing any $1$-edge with a $1$-path of arbitrary length; $\cF_k$ may be seen as the set of successors of $\cA_k$.
At this point, to give a complete description of $\cF_k$, it only remains to retrieve all the elements of $\cA_k$.

\begin{lemma} \label{lem:k_bound}
    Let $\bm{x}$ be an ancestor in $\cA_k$ and let $n$ be its number of nodes.
    Then \ $k + 3 \leq n \leq 2k$.
\end{lemma}

Therefore, the elements of $\cA_k$ are finite and can be recovered from the lists of vertices of $\psep{n}$ with $n$ up to $2k$ (provided we have them at our disposal).
We simply need to scroll through the lists of vertices of $\psep{n}$ for all the $n = k \!+\! 3, \dots, 2k$, and extract the ones with exactly $n + k$ edges and no node of degree $2$.
Since, at the time being, an exhaustive list of fractional vertices (up to isomorphism%
\footnote{An isomorphism between vertices is a relabeling of their nodes.})
is already available for $n$ up to $12$ \cite{art:BenBoy:IGSmallTSP,art:BoyEll:ExtrSEP}, it is possible to completely determine $\cA_3, \cA_4, \cA_5, \cA_6$ (up to isomorphism) by \Cref{lem:k_bound}. 
For instance, \Cref{fig:vertices_A4} shows all the ancestors in $\cA_4$: as $k = 4$, they can be extracted from the lists of vertices of $\psep{7}$ and $\psep{8}$.
All the vertices of $\cF_4$ are either of these shapes or with the $1$-edges replaced with $1$-paths of arbitrary length.
Similarly, simply examining the structure of the sole ancestor in $\cA_3$, one can deduce that the family $\cF_3$ consists precisely of those vertices formed by two $\nicefrac{1}{2}$-triangles connected by three $1$-paths.
In virtue of this characterization, one may see that Conjecture~4.1 in~\cite{art:BenBoy:IGSmallTSP} supports the Few Edge Hypothesis.

\begin{figure}
    \centering
    \newcommand{\preparecubecoordinates}{
    \node at (0,   0.5) (a) {};
    \node at (1.5, 0.5) (b) {};
    \node at (1.5, 2)   (c) {};
    \node at (0,   2)   (d) {};
    \node at (2.5, 0)   (e) {};
    \node at (4,   0)   (f) {};
    \node at (4,   1.5) (g) {};
    \node at (2.5, 1.5) (h) {};
}

\begin{tikzpicture}[x=5mm, y=5mm]
    \begin{scope}[shift={(0, 0)}]
        \node at (0, 0) (a) {};   \node at (4, 0) (e) {};
        \node at (2, 0) (b) {};   \node at (4, 2) (f) {};
        \node at (0, 2) (c) {};   \node at (3, 1) (g) {};
        \node at (2, 2) (d) {};
        \draw plot coordinates {(a) (g) (e) (b) (d) (f) (g) (c) (a)};
        \draw[double] plot coordinates {(a) (b)}
                      plot coordinates {(c) (d)}
                      plot coordinates {(e) (f)};
        \foreach \vname in {a, b, c, d, e, f, g}{
            \fill (\vname) circle (2pt);
        }
    \end{scope}
    
    \begin{scope}[shift={(5, 0)}]
        \preparecubecoordinates
        \draw plot coordinates {(a) (e) (f) (b) (c) (g) (h) (d) (a)};
        \draw[double] plot coordinates {(a) (b)}
                      plot coordinates {(c) (d)}
                      plot coordinates {(e) (h)}
                      plot coordinates {(f) (g)};
        \foreach \vname in {a, b, c, d, e, f, g, h}{
            \fill (\vname) circle (2pt);
        }
    \end{scope}
    
    \begin{scope}[shift={(10, 0)}]
        \preparecubecoordinates
        \draw plot coordinates {(a) (e) (d) (a)};
        \draw plot coordinates {(f) (b) (c) (g) (h) (f)};
        \draw[double] plot coordinates {(a) (b)}
                      plot coordinates {(c) (d)}
                      plot coordinates {(e) (h)}
                      plot coordinates {(f) (g)};
        \foreach \vname in {a, b, c, d, e, f, g, h}{
            \fill (\vname) circle (2pt);
        }
    \end{scope}
    
    \begin{scope}[shift={(15, 0)}]
        \preparecubecoordinates
        \draw[double] plot coordinates {(a) (d)}
                      plot coordinates {(b) (c)};
        \draw plot coordinates {(a) (e) (g) (c) (d) (h) (f) (b) (a)};
        \draw[double] plot coordinates {(e) (h)}
                      plot coordinates {(f) (g)};
        \foreach \vname in {a, b, c, d, e, f, g, h}{
            \fill (\vname) circle (2pt);
        }
    \end{scope}
    
    \begin{scope}[shift={(20, 0)}]
        \preparecubecoordinates
        \draw plot coordinates {(a) (e) (g) (c) (b) (f) (h) (d) (a)};
        \draw[double] plot coordinates {(a) (b)}
                      plot coordinates {(c) (d)}
                      plot coordinates {(e) (h)}
                      plot coordinates {(f) (g)};
        \foreach \vname in {a, b, c, d, e, f, g, h}{
            \fill (\vname) circle (2pt);
        }
    \end{scope}
\end{tikzpicture}

\vspace{0.3cm}
\begin{tikzpicture}[x=5mm, y=5mm]
    \draw[double] (0, 0) -- (1.5, 0) node[right] {$1$};
    \draw         (3, 0) -- (4.5, 0) node[right] {$\frac{1}{2}$};
\end{tikzpicture}
    \caption{Vertices of $\cA_4$, up to isomorphism.}
    \label{fig:vertices_A4}
\end{figure}

%%%%%%%%%%%%%%%%%%%%%%%%%
% Redefining the Gap problem
%%%%%%%%%%%%%%%%%%%%%%%%%
\section{Redefining the \texorpdfstring{\gap}{Gap} problem} \label{sec:gapp}

In this section, we formulate a relaxed definition of \gap.

\begin{definition}[\gapp]
    Let $\bm{x} \in \psep{n}$ be a vertex. The \emph{\gapp\ of $\bm{x}$} is
    $$
    \gapp(\bm{x}) := \sup \left\{ \frac{\tsp(\bm{c})}{\bm{c} \bm{x}} \ \mid \ \bm{c} \text{ metric } \right\} \ .
    $$
\end{definition}

By removing the requirement $\bm{x} \in \arg\min \sep(\bm{c})$, we obtain the inequality:
\begin{equation} \label{eq:gapp_vs_gap}
    \gapp(\bm{x}) \geq \gap(\bm{x}) \ .
\end{equation}
Our objective is therefore to provide an upper bound on \gapp, which consequently yields a valid upper bound for \gap.
To compute the inverse of $\gap(\bm{x})$, Benoit and Boyd, in~\cite{art:BenBoy:IGSmallTSP}, designed an LP named $\opt(\bm{x})$.
Adopting the same strategy, for a given a vertex $\bm{x}$, we consider a corresponding linear problem $\optp(\bm{x})$, defined as follows:
{\small
\begin{align}
    \text{minimize \ \ }
        & \sum_{ij \in E_n} x_{ij} c_{ij}
            \label{eq:optp_obj} \\
    \text{subject to: \ }
        & c_{ik} + c_{jk} - c_{ij} \ \geq 0
            & & \forall ij \in E_n, \ k \neq i, j,
            \label{eq:optp_tr_ineq} \\
        & \sum_{ij\in E_n} t_{ij} c_{ij} \ \geq 1
            & & \forall \bm{t} \text{ tour},
            \label{eq:optp_tour_cost_1} \\
        & c_{ij} \geq 0
            & & \forall ij \in E_n.
            \label{eq:optp_var}
\end{align}}
Constraints~\eqref{eq:optp_tr_ineq} force $\bm{c}$ to be a metric cost;
constraints~\eqref{eq:optp_tour_cost_1} normalize $\tsp(\bm{c}) = 1$;
minimizing $\bm{c} \bm{x}$ is then equivalent to maximizing $1 / \bm{c} \bm{x} = \tsp(\bm{c}) / \bm{c} \bm{x}$, \ thus
\begin{equation} \label{eq:optp_and_gapp}
    \frac{1}{\optp(\bm{x})} = \gapp(\bm{x}) \ .
\end{equation}

The following lemma is the first result that studies the behavior of \gapp\ under the application of the \bb-move.

\begin{lemma} \label{lem:bb_move_gap}
    Let $\bm{x} \in \psep{n}$.
    Then $\gapp(\, \bb(\bm{x}) \, ) \geq \gapp(\bm{x})$.
    That is, the \bb-move is \gapp-increasing.
\end{lemma}

%%%%%%%%%%%%%%%%%%%%%%%%%
% The Gap-Bounding algorithm
%%%%%%%%%%%%%%%%%%%%%%%%%
\section{The \texorpdfstring{\gap}{Gap}-Bounding algorithm} \label{sec:gap_bounding}

This section is entirely devoted to finding a way to bound the \gapp\ for all the vertices of a given family $\cF_k$.
To this end, we aim to contain the increase in \gapp\ originated by an iterative application of a \bb-move.

We begin by outlining the central idea of our approach.
Our goal is to provide a lower bound on \optp, which results in an upper bound on \gapp\ by \eqref{eq:optp_and_gapp}.
Assume that we are given a vertex $\bm{x}_0$ alongside with an optimal solution of $\optp(\bm{x}_0)$ and a corresponding optimal solution $\bm{\mu}^0$ of the dual problem $\dual\optp(\bm{x}_0)$.
When considering a successor $\bm{x}'$ of $\bm{x}_0$, we intend to construct a feasible dual solution $\bm{\mu}'$ for $\dual\optp(\bm{x}')$ starting from $\bm{\mu}^0$: succeeding in this task will directly result in a lower bound for $\optp(\bm{x}')$, as guaranteed by duality theory.

Indeed, for every vertex $\bm{x}$, equations \eqref{eq:gapp_vs_gap}, \eqref{eq:optp_and_gapp}, and duality theory yield
\begin{equation} \label{eq:chain_ineq}
    \gap(\bm{x}) \leq \gapp(\bm{x})
    = \frac{1}{\optp(\bm{x})}
    = \frac{1}{\dual\optp(\bm{x})}
    \leq \frac{1}{L} \ ,
\end{equation}
where $L$ is any lower bound on $\dual\optp(\bm{x})$.
In our case, $L$ will be a bound on the objective value attained by $\bm{\mu}'$ (elaborated from $\bm{\mu}^0$).
This strategy is visually represented in \Cref{fig:flowchart_tech}: it will be the skeleton of the \gap-Bounding algorithm, discussed in detail in \Cref{subsec:bounding_gap}.

\begin{figure}
    \centering
    \begin{tikzpicture}[x=1.3cm, linebreak/.style={align=center}]
    \foreach \x in {0, 2, 4, 6, 8}{
        \draw (\x, 0) circle (4mm);
    }
    \foreach \x in {0, 2, 4, 6}{
        \draw[->, dashed] (\x + 0.5, 0) -- (\x + 1.5, 0);
    }
    \node at (0, 0) {$\bm{x}_0$};
    \node at (2, 0) {$\bm{\mu}^0$};
    \node at (4, 0) {$\bm{\mu}'$};
    \node at (6, 0) {$L$};
    \node at (8, 0) {$\frac{1}{L}$};

    \node at (0, 0.75) {\textsc{input:}};
    \node at (8, 0.75) {\textsc{output:}};

    \node[below, linebreak] at (0, -0.5) {vertex};
    \node[below, linebreak] at (2, -0.5) {opt sol of \\ $\dual\optp(\bm{x}_0)$};
    \node[below, linebreak] at (4, -0.5) {feas sol of \\ $\dual\optp(\bm{x}')$};
    \node[below, linebreak] at (6, -0.5) {lower bound on \\ the obj value of $\bm{\mu}'$};
    \node[below, linebreak] at (8, -0.5) {$\gap(\bm{x}') \leq \frac{1}{L}$ \\ \eqref{eq:chain_ineq}};
\end{tikzpicture}
    \caption{Flowchart of the pipeline applied to bound \gap\ for a successor $\bm{x}'$ of $\bm{x}_0$.}
    \label{fig:flowchart_tech}
\end{figure}

Clearly, the crucial point in this procedure is to find a ``good-enough'' feasible dual solution, so that the bound obtained is meaningful (e.g., the trivial dual solution of all zeros ultimately gives no bound on \gapp).

%%%%%%%%%% Dual formulations of the OPT+ problem
\subsection{Dual formulations of the \texorpdfstring{\optp}{OPT+} problem} \label{sec:doptp_ii}

Given a vertex $\bm{x}$, the dual problem $\dual\optp(\bm{x})$ is derived as follows.
{\small
\begin{align}
    \text{maximize \ \ }
        & \sum_{\bm{t} \text{ tour }} \mu_{\bm{t}}
            \label{eq:doptp_obj} \\
    \text{subject to: \ }
        & \sum_{k \neq i, j}(-\lambda_{ijk} +\lambda_{ikj} +\lambda_{jki})
            + \!\! \sum_{\bm{t} \text{ tour }} t_{ij} \mu_{\bm{t}}
            \ \leq x_{ij}
            & & \forall ij \in E_n,
            \label{eq:doptp_constr} \\
        & \lambda_{ijk} \ \geq 0
            & & \forall ij \in E_n, \ k \neq i, j,
            \label{eq:doptp_var_lambda} \\
        & \mu_{\bm{t}} \ \geq 0
            & & \forall \bm{t} \text{ tour on $K_n$}.
            \label{eq:doptp_var_mu}
\end{align}}
Notice that the abuse of notation $ij \equiv ji \in E$ induces also $\lambda_{ijk} \equiv \lambda_{jik}$; this fact does not involve the third index: $\lambda_{ijk} \not\equiv \lambda_{kji}$
\footnote{With the symbol $\equiv$, we do not mean that two distinct variables have the same value; rather, we mean that the two notations coincide: they denote the same variable.}.

When considering a vertex $\bm{x}$ and one of its successors $\bm{x}'$, the task of producing a feasible solution of $\dual\optp(\bm{x}')$ starting from an optimal solution of $\dual\optp(\bm{x})$ is anything but trivial.
In what follows, we design an equivalent formulation and make use of it instead: \ $\dual\optii$.
Its main purpose is to get rid of $\bm{\lambda}$ variables: this is achieved at the price of considering walks on $G_{\bm{x}}$ (see \Cref{def:walk} and \Cref{rem:walk_mult}) in place of tours.
For a given vertex $\bm{x}$, $\dual\optii(\bm{x})$ is thus defined as:
{\small
\begin{align}
    \text{maximize \ \ }
        & \sum_{\bm{w} \text{ walk on $G_{\bm{x}}$}} \mu_{\bm{w}}
            \label{eq:doptii_obj} \\
    \text{subject to: \ }
        & \sum_{\bm{w} \text{ walk on $G_{\bm{x}}$}} w_{ij} \mu_{\bm{w}} \ \leq x_{ij}
            & & \forall ij \in E_{\bm{x}},
            \label{eq:doptii_constr} \\
        & \mu_{\bm{w}} \ \geq 0
            & & \forall \bm{w} \text{ walk on $G_{\bm{x}}$}.
            \label{eq:doptii_var_mu}
\end{align}}

The equivalence of the two formulations $\dual\optp(\bm{x})$ and $\dual\optii(\bm{x})$ (that is, they have the same optimal value) is stated in the following lemma.

\begin{lemma} \label{lem:doptp_ii}
    $\dual\optp(\bm{x}) = \dual\optii(\bm{x})$ \ for every vertex $\bm{x}$ of $\psep{n}$.
\end{lemma}

%%%%%%%%%% Bounding the gap on successors
\subsection{Bounding the \texorpdfstring{\gap}{Gap} on successors} \label{subsec:bounding_gap}

The goal of this section is to derive, for a generic successor $\bm{x}'$ of $\bm{x}_0$, a feasible solution of $\dual\optii(\bm{x}')$, starting from an optimal solution of $\dual\optii(\bm{x}_0)$; this serves as a bound on $\gapp(\bm{x}')$
(see \Cref{fig:flowchart_tech}, with $\dual\optii$ in place of $\dual\optp$).

Let $\bm{x}_0$ be a vertex of $\psep{n}$ and let $\bm{\mu}^0 \in \arg\max \dual\optii(\bm{x}_0)$.
We begin by considering a $1$-edge $ab$ of $\bm{x}_0$ and applying $d$ consecutive \bb-moves (see \Cref{def:BBmove}): we insert nodes \ $a_1, \dots, a_d$ \ and obtain, by \Cref{thm:bb_move}, the vertices \ $\bm{x}_k := \bb(\bm{x}_{k-1}, \ a_{k-1} b)$ \ for $k = 1, \dots, d$ \ (where $a_0 := a$).
We aim to design a feasible solution for 
$\dual\optii(\bm{x}_d)$.

An assignment $\bm{\mu}^d$ for $\dual\optii(\bm{x}_d)$ is indexed by the walks on $G_{\bm{x}_d}$.
We construct walks $\bm{w}^d$ on $G_{\bm{x}_d}$, starting from the ones on $G_{\bm{x}_0}$, and assign weights to the corresponding components of $\bm{\mu}^d$.
In this perspective, we divide the walks on $G_{\bm{x}_0}$ into three families, according to how many times the edge $ab$ is traversed: \
we define \ $\cW^{ab}_m := \{ \bm{w} \text{ walk} \ \mid \ w_{ab} = m \}$ \ for $m = 0, 1, 2$.

If $\bm{w}^0 \in \cW^{ab}_0$, we construct the following $d \!+\! 1$ walks on $G_{\bm{x}_d}$. \
For each $k = 0, \dots, d$, \ the walk $\bm{w}^d_k$ retraces $\bm{w}^0$ but, when it arrives at $a$, it deviates to pass through $a_1, a_2, \dots, a_k$ and back, and when it arrives at $b$, it deviates to $a_d, a_{d-1}, \dots, a_{k+1}$ and back.
We set $\mu^d_{\bm{w}^d_k} := \frac{1}{d+1} \mu^0_{\bm{w}^0}$ \ for all $k = 0, \dots, d$.

If $\bm{w}^0 \in \cW^{ab}_1$, we construct the walk $\bm{w}^d$ on $G_{\bm{x}_d}$ that retraces $\bm{w}^0$ but replaces $ab$ with $a a_1, a_1 a_2, \dots, a_{d-1} a_d, a_d b$.
We set $\mu^d_{\bm{w}^d} := \mu^0_{\bm{w}^0}$.

If $\bm{w}^0 \in \cW^{ab}_2$, we construct the walk $\bm{w}^d$ on $G_{\bm{x}_d}$ that retraces $\bm{w}^0$ but replaces the double passage on $ab$ by passing twice through $a a_1, a_1 a_2, \dots, a_{d-1} a_d, a_d b$.
We set $\mu^d_{\bm{w}^d} := \mu^0_{\bm{w}^0}$.

Notice that these constructions cover all the possible walks $\bm{w}^d$ on $G_{\bm{x}_d}$.
\Cref{fig:BB_walks_constructions} illustrates the new assignment $\bm{\mu}^d$ for $d = 2$.

\begin{figure}[htbp]
    \centering
    \newcommand{\preparexzerocoordinates}{
    \node at (0,    0)    (a)    {};
    \node at (1.5,  0)    (b)    {};
    \node at (-1,   +0.5) (p1)   {};
    \node at (-1,   -0.5) (p2)   {};
    \node at (+2.5, +0.5) (q1)   {};
    \node at (+2.5, -0.5) (q2)   {};
    \node at (0,    2pt)  (epsy) {};
}
\newcommand{\preparextwocoordinates}{
    \node at (0,  0)    (a)    {};
    \node at (1,  0)    (a_1)  {};
    \node at (2,  0)    (a_2)  {};
    \node at (3,  0)    (b)    {};
    \node at (-1, +0.5) (p1)   {};
    \node at (-1, -0.5) (p2)   {};
    \node at (+4, +0.5) (q1)   {};
    \node at (+4, -0.5) (q2)   {};
    \node at (0,  2pt)  (epsy) {};
}
\newcommand{\drawdoublepassage}[3]{
    \draw[#3] plot coordinates {($(#1) + (epsy)$) ($(#2) + (epsy)$)};
    \draw[#3] plot coordinates {($(#1) - (epsy)$) ($(#2) - (epsy)$)};
}
\newcommand{\drawbackandforth}[3]{
    \draw[#3] plot coordinates {($(#1) + (0, +2pt)$) (#2) ($(#1) + (0, -2pt)$)};
}

\begin{tikzpicture}[scale=0.6]
    \begin{scope}[xshift=-7cm]
        \node at (0, 0)      {$\bm{w}^0 \in \cW^{ab}_0$};
        \begin{scope}[shift={(-0.75, -1.5)}]
            \preparexzerocoordinates
            \foreach \vname in {a, b}{
                \fill (\vname) circle (2pt);
                \node[above=3pt] at (\vname) {$\vname$};
            }
            \draw[violet] plot coordinates {(p1) (a) (p2)}
                          plot coordinates {(q1) (b) (q2)};
            \node at (-1.25, -0.25) {\small $\bm{w}^0$};
        \end{scope}
        \node[below] at (0, -1.8) {$\mu^0_{\bm{w}^0}$};
        
        \draw[|->] (0, -3) -- (0, -3.5);
        
        \begin{scope}[shift={(-1.5, -4.5)}]
            \node at (-1.25, -0.25) {\small $\bm{w}^2_0$};
            \preparextwocoordinates
            \foreach \vname in {a, a_1, a_2, b}{
                \fill (\vname) circle (2pt);
                \node[above=3pt] at (\vname) {$\vname$};
            }
            
            \draw[violet] plot coordinates {(p1) (a) (p2)};
            \draw[violet] plot coordinates {(q1) ($(b) + (epsy)$)}
                          plot coordinates {(q2) ($(b) - (epsy)$)};
            \drawdoublepassage{a_2}{b}{violet}
            \drawbackandforth{a_2}{a_1}{violet}
        \end{scope}
        
        \begin{scope}[shift={(-1.5, -6)}]
            \node at (-1.25, -0.25) {\small $\bm{w}^2_1$};
            \preparextwocoordinates
            \foreach \vname in {a, a_1, a_2, b}{
                \fill (\vname) circle (2pt);
                \node[above=3pt] at (\vname) {$\vname$};
            }
            
            \draw[violet] plot coordinates {(p1) ($(a) + (epsy)$)}
                          plot coordinates {(p2) ($(a) - (epsy)$)};
            \draw[violet] plot coordinates {(q1) ($(b) + (epsy)$)}
                          plot coordinates {(q2) ($(b) - (epsy)$)};
            \drawbackandforth{a}{a_1}{violet}
            \drawbackandforth{b}{a_2}{violet}
        \end{scope}
        
        \begin{scope}[shift={(-1.5, -7.5)}]
            \node at (-1.25, -0.25) {\small $\bm{w}^2_2$};
            \preparextwocoordinates
            \foreach \vname in {a, a_1, a_2, b}{
                \fill (\vname) circle (2pt);
                \node[above=3pt] at (\vname) {$\vname$};
            }
            
            \draw[violet] plot coordinates {(p1) ($(a) + (epsy)$)}
                          plot coordinates {(p2) ($(a) - (epsy)$)};
            \draw[violet] plot coordinates {(q1) (b) (q2)};
            \drawdoublepassage{a}{a_1}{violet}
            \drawbackandforth{a_1}{a_2}{violet}
        \end{scope}
        \node[below] at (0, -8) {$\mu^2_{\bm{w}^2_k} := \frac{1}{3} \ \mu^0_{\bm{w}^0}$};
    \end{scope}

    \begin{scope}
        \node at (0, 0)      {$\bm{w}^0 \in \cW^{ab}_1$};
        \begin{scope}[shift={(-0.75, -1.5)}]
            \preparexzerocoordinates
            \foreach \vname in {a, b}{
                \fill (\vname) circle (2pt);
                \node[above=3pt] at (\vname) {$\vname$};
            }
            \draw[blue] plot coordinates {(p2) (a) (b) (q2)};
            \node at (-1.25, -0.25) {\small $\bm{w}^0$};
        \end{scope}
        \node[below] at (0, -1.8) {$\mu^0_{\bm{w}^0}$};
        
        \draw[|->] (0, -3) -- (0, -3.5);
        
        \begin{scope}[shift={(-1.5, -4.5)}]
            \node at (-1.25, -0.25) {\small $\bm{w}^2$};
            \preparextwocoordinates
            \foreach \vname in {a, a_1, a_2, b}{
                \fill (\vname) circle (2pt);
                \node[above=3pt] at (\vname) {$\vname$};
            }
            
            \draw[blue] plot coordinates {(p2) (a) (a_1) (a_2) (b) (q2)};
        \end{scope}
        \node[below] at (0, -5) {$\mu^2_{\bm{w}^2} := \mu^0_{\bm{w}^0}$};
    \end{scope}
    
    \begin{scope}[xshift=+7cm]
        \node at (0, 0)      {$\bm{w}^0 \in \cW^{ab}_1$};
        \begin{scope}[shift={(-0.75, -1.5)}]
            \preparexzerocoordinates
            \foreach \vname in {a, b}{
                \fill (\vname) circle (2pt);
                \node[above=3pt] at (\vname) {$\vname$};
            }
            \draw[cyan] plot coordinates {(p1) ($(a) + (epsy)$)}
                        plot coordinates {(p2) ($(a) - (epsy)$)};
            \drawdoublepassage{a}{b}{cyan}
            \draw[cyan] plot coordinates {(q1) ($(b) + (epsy)$)}
                        plot coordinates {(q2) ($(b) - (epsy)$)};
            \node at (-1.25, -0.25) {\small $\bm{w}^0$};
        \end{scope}
        \node[below] at (0, -1.8) {$\mu^0_{\bm{w}^0}$};
        
        \draw[|->] (0, -3) -- (0, -3.5);
        
        \begin{scope}[shift={(-1.5, -4.5)}]
            \node at (-1.25, -0.25) {\small $\bm{w}^2$};
            \preparextwocoordinates
            \foreach \vname in {a, a_1, a_2, b}{
                \fill (\vname) circle (2pt);
                \node[above=3pt] at (\vname) {$\vname$};
            }
            
            \draw[cyan] plot coordinates {(p1) ($(a) + (epsy)$)}
                        plot coordinates {(p2) ($(a) - (epsy)$)};
            \drawdoublepassage{a}{a_1}{cyan}
            \drawdoublepassage{a_1}{a_2}{cyan}
            \drawdoublepassage{a_2}{b}{cyan}
            \draw[cyan] plot coordinates {(q1) ($(b) + (epsy)$)}
                        plot coordinates {(q2) ($(b) - (epsy)$)};
        \end{scope}
        \node[below] at (0, -5) {$\mu^2_{\bm{w}^2} := \mu^0_{\bm{w}^0}$};
    \end{scope}

\end{tikzpicture}
    \caption{Variables $\bm{\mu}^2$ for $\dual\optii(\bm{x}_2)$ \ constructed from variables $\bm{\mu}^0$ for $\dual\optii(\bm{x}_0)$.}
    \label{fig:BB_walks_constructions}
\end{figure}

In order to extend the construction for a generic successor of $\bm{x}_0$, we sequentially repeat the previous procedure for each $1$-edge $e_1, \dots, e_p$ of $\bm{x}_0$, expanding them into $1$-paths with $d_1, \dots, d_p$ internal nodes.
This yields an assignment $\bm{\mu}'$ for $\dual\optii(\bm{x}')$ such that: \
(i) $\bm{\mu}'$ attains the same objective value as $\bm{\mu}^0$; \
(ii) $\bm{\mu}'$ satisfies all the constraints~\eqref{eq:doptii_constr}, except those associated to the $1$-edges, which nonetheless remain bounded by fixed constants.
These constants are the same for all the edges in the same $1$-path of $\bm{x}'$, originated from the same $1$-edge $e_h$ of $\bm{x}_0$:
$$
C(\bm{x}_0, e_h) :=
2 \!\!\! \sum_{\bm{w}^0 \in \cW^{e_h}_0} \mu^0_{\bm{w}^0} \ + \
         \sum_{\bm{w}^0 \in \cW^{e_h}_1} \mu^0_{\bm{w}^0} \ + \
2 \!\!\! \sum_{\bm{w}^0 \in \cW^{e_h}_2} \mu^0_{\bm{w}^0} \ .
$$
Crucially, they only depend on the initial vertex $\bm{x}_0$, not on the lengths of the $1$-paths of $\bm{x}'$.
Rescaling by the factor \
$C^*(\bm{x}^0) := \max_{h = 1, \dots, p} \{ C(\bm{x}_0, e_h) \}$, \
we thus obtain the feasible solution \
$\bm{\mu}^* := \frac{1}{C^*(\bm{x}^0)} \bm{\mu}'$, \
whose objective value accordingly becomes $\frac{1}{C^*(\bm{x}^0)} \dual\optii(\bm{x}_0)$.
This value plays the role of $L$ in \cref{eq:chain_ineq}, yielding \
$\gapp(\bm{x}') \ \leq \ C^*(\bm{x}^0) \cdot \frac{1}{\dual\optii(\bm{x}_0)}$, \
and ultimately, by eq.~\eqref{eq:optp_and_gapp} and  \Cref{lem:doptp_ii},
$$
\gapp(\bm{x}') \ \leq \ C^*(\bm{x}^0) \cdot \gapp(\bm{x}_0) \ .
$$

Notice that we expect $C^*(\bm{x}^0)$ to be greater than or equal to $1$, otherwise we would obtain $\gapp(\bm{x}_d) < \gapp(\bm{x}_0)$, contradicting \Cref{lem:bb_move_gap}.

We are finally able to design the \gap-Bounding (\gb) algorithm as \Cref{alg:GB_algo}; its main purpose is clarified in the next theorem.

\begin{algorithm}
    \caption{\gap-Bounding algorithm} \label{alg:GB_algo}
    \begin{algorithmic}[1]
        \STATE \textsc{input}: $\bm{x} \in \psep{n}$ vertex \\[5pt]
        \STATE Solve $\dual\optii(\bm{x})$ and retrieve an optimal assignment of variables $\{ \mu_{\bm{w}} \}_{\bm{w} \text{ walk on } \bm{x}}$.
        \FOR{each $e$ $1$-edge of $\bm{x}$}
            \STATE $\displaystyle
                C(\bm{x}, e) \ \gets \
                2 \sum_{\bm{w} \in \cW^e_0} \mu_{\bm{w}} +
                \sum_{\bm{w} \in \cW^e_1} \mu_{\bm{w}} +
                2 \sum_{\bm{w} \in \cW^e_2} \mu_{\bm{w}}$
        \ENDFOR
        \STATE $C^*(\bm{x}) \ \gets \ \max \{ C(\bm{x}, e) \}_{e \text{ $1$-edge of } \bm{x}}$
        \STATE $\gapp(\bm{x}) \ \gets \ \nicefrac{1}{\dual\optii(\bm{x})}$
        \STATE Return: $C^*(\bm{x}) \cdot \gapp(\bm{x})$
    \end{algorithmic}
\end{algorithm}

\begin{theorem} \label{thm:GB_algo}
    The \gap-Bounding algorithm, on input a vertex $\bm{x} \in \psep{n}$, returns a value $\gb(\bm{x})$ which is an upper bound for the gap of all the successors $\bm{x}'$ of $\bm{x}$: \
    $\gap(\bm{x}') \leq \gb(\bm{x})$.
\end{theorem}

\begin{proof}
    This whole section is the proof that \ $\gapp(\bm{x}') \leq C^*(\bm{x}) \cdot \gapp(\bm{x}) = \gb(\bm{x})$ \ for any successor $\bm{x}'$ of $\bm{x}$; equation~\eqref{eq:gapp_vs_gap} completes the argument.
    \qed
\end{proof}

Notice that both the \gb\ algorithm and \Cref{thm:GB_algo} apply to all generic $\bm{x}$ vertices, without requiring that they be ancestors.
The following lemma shows how to exploit this fact.

\begin{lemma}\label{lem:bound_ancestors}
    Let $\bm{x} \in \psep{n}$ be a vertex and let $\bm{x}_0$ be its ancestor.
    Then $\gb(\bm{x})$ gives a bound for the gap of all the successors $\bm{x}'$ of $\bm{x}_0$ (even for $\bm{x}'$ that are not successors of $\bm{x}$): \
    $\gap(\bm{x}') \leq \gb(\bm{x})$.
\end{lemma}

%%%%%%%%%%%%%%%%%%%%%%%%%
% Computational results
%%%%%%%%%%%%%%%%%%%%%%%%%
\section{Computational results} \label{sec:comp_results}

This section reports how the execution of the \gap-bounding algorithm yields the \emph{computer-aided} proof of our main result: \Cref{thm:Gap_4over3}.

Our pipeline starts by extracting the ancestors of $\cA_k$ for $k = 3, 4, 5, 6$ from the lists of vertices already available from \cite{art:BenBoy:IGSmallTSP,art:BoyEll:ExtrSEP}%
\footnote{Vertices data can be found at this \href{www.site.uottawa.ca/~sylvia/subtourvertices/index.htm}{link}, last visited 04.06.2025.}, as explained in \Cref{sec:families}.
Subsequently, the \gap-Bounding algorithm is applied on each vertex $\bm{x} \in \cA_k$.

We inform that, for some ancestors, a straightforward application of the \gb\ algorithm returned a value higher than $\nicefrac{4}{3}$.
To improve the bounds returned, we exploited \Cref{lem:bound_ancestors} and executed the \gb\ algorithm on some selected successors.
Even though there is no guarantee that the \gb\ algorithm applied to successors gives strictly better results, we thought it reasonable that more ``information'' (more nodes, edges, and dual variables) would lead to a more detailed analysis and a tighter bound.
Eventually, this intuition led us to accomplish our objective.

\begin{theorem} \label{thm:Gap_4over3}
    $\gap(\bm{x}) \leq \frac{4}{3}$ \ for all the vertices of $\bm{x} \in \cF_k$ \ with $k = 3, 4, 5, 6$.
\end{theorem}

We implemented the \gb\ algorithm in Python, employing the commercial software Gurobi \cite{gurobi} to solve the LPs.
Our code is available \href{https://github.com/tulliovilla/IG_for_TSP_with_few_edges}{here}.
The computation is relatively lightweight and can be easily performed on a standard laptop.

%%%%%%%%%%%%%%%%%%%%%%%%%
% Conclusion
%%%%%%%%%%%%%%%%%%%%%%%%%
\section{Conclusion} \label{sec:conclusion}

In this paper, the main result of proving the $\nicefrac{4}{3}$-conjecture for vertices of $\psep{n}$ with at most $n + 6$ edges in their support graphs is presented.
A natural continuation of this research consists of constructing $\cA_7$ by exhaustive enumeration, to enlarge the set of ancestors on which to apply the \gb\ algorithm and collect bounds for the whole family $\cF_7$.

Besides this, the new \emph{methodology} introduced is itself of independent interest, as it opens a new range of possible advancements in the field.
Investigating novel transformations between vertices and analyzing their impact on the integrality gap (akin to the analysis we presented for the \bb-move) may lead to interesting results on the integrality gap for diverse infinite families of vertices.

To conclude, we also mention a last research direction, motivated by an intriguing, computationally verified observation.
Using data from \cite{art:BenBoy:IGSmallTSP}, for instances with few nodes, we noticed a correlation between the number of edges in the support graphs of the vertices of $\psep{n}$ (with $6 \leq n \leq 12$) and their integrality gap.
As shown in \Cref{fig:gap_vs_n_edges}, for a fixed $n \leq 12$, the integrality gap is higher on vertices with few non-zero components and appears to decrease as the number of edges in the support graph increases.
Although this correlation is currently supported only for small values of $n$, it presents a fascinating avenue for further investigation.
For example, the donut instances of \cite{boyd2021salesman} have an integrality gap tending to $\nicefrac{4}{3}$, and they do not belong to the families of instances considered in our result; nevertheless, for each donut instance, the proposed lower bound on its integrality gap is strictly smaller than the integrality gap of a vertex of $\cF_3$ with the same number of nodes.
Should one succeed in proving that, for every $n$, the maximum integrality gap on the polytope $\psep{n}$ is always attained on the vertices with the smallest number of edges, such a result together with the contribution of this work would be the two key ingredients to definitely resolve the $\nicefrac{4}{3}$-conjecture.

\begin{figure}
    \centering
    \input{tikz_plots/gap_vs_n_edges}
    \caption{Correlation between the number of edges in the support graph of a vertex and their integrality gap.
    Each plot collects all the vertices of $\psep{n}$ for a fixed $n$: \ $n = 10$ on the left, \ $n = 11$  on the right.
    The observed trend is the same for all small $n \leq 12$.}
    \label{fig:gap_vs_n_edges}
\end{figure}

%%%%%%%%%%%%%%%%%%%%%%%%%
% Acknowledgments
%%%%%%%%%%%%%%%%%%%%%%%%%
\section*{Acknowledgments}

This work was supported by the Swiss National Science Foundation project
n. 200021\_212929/1 ``Computational methods for integrality gaps analysis''.

%%%%%%%%%%%%%%%%%%%%%%%%%
% Bibliography
\bibliographystyle{splncs04}
\bibliography{bibliography}

%%%%%%%%%%%%%%%%%%%%
% Appendix
%%%%%%%%%%%%%%%%%%%%
\appendix

%%%%%%%%%%%%%%%%%%%%
% Auxiliary proofs
%%%%%%%%%%%%%%%%%%%%
\section{Auxiliary proofs} \label{sec:aux_proofs}

In this appendix, we discuss in detail all the proofs that have been removed from the main argument exposition.
The self-contained proofs are presented consecutively, while a separate section is devoted to the more intricate arguments that warrant further discussion.

\begin{proof}[of \Cref{lem:min_edges}]
    Consider three distinct $1$-paths of $\bm{x}$ (they exist by \Cref{thm:3_1paths}) and let $\tilde{V}$ be the set of their end nodes, so that \ $|\tilde{V}| = 6$ \ and \ $\deg(v) \geq 3 \ \ \forall v \in \tilde{V}$. \
    Then \
    $|E_{\bm{x}}|
    = \frac{1}{2} \sum_{v \in V_n} \deg(v)
    = \frac{1}{2} \big( \sum_{v \in \tilde{V}} \deg(v)
    + \sum_{v \notin \tilde{V}} \deg(v) \big)
    \geq \frac{1}{2} \big( 6 \!\cdot\! 3 + (n-6) \!\cdot\! 2 \big)
    = n + 3$.
    \qed
\end{proof}

\begin{proof}[of \Cref{lem:k_bound}]
    Theorem~\ref{thm:max_edges} gives the inequality \ $n + k = |E_{\bm{x}}| \leq 2n - 3$ \ which leads to $k + 3 \leq n$.
    The upper bound on $n$ is derived from the fact that the minimum degree of the nodes of any ancestor $\bm{x}$ is $3$, thus \
    $n + k = |E_{\bm{x}}| = \frac{1}{2} \sum_{v \in V_n} \deg(v) \geq \frac{1}{2} \cdot 3 \, n$ \
    which simplifies to \ $n \leq 2k$.
    \qed
\end{proof}

\begin{proof}[of \Cref{lem:bb_move_gap}]
    Let $\bm{x}_0 \in \psep{n}$ be a vertex with a $1$-edge $ab$.
    Let $\bm{x}_1 \in \psep{n+1}$ be the result of the \bb-move applied on $ab$: we denote $w$ the node added and $aw, wb$ the two $1$-edges originated by this move.
    Consider a metric cost $\bm{c}^0$ which realizes the \gapp\ on $\bm{x}_0$, namely
    $\gapp(\bm{x}_0) = \frac{\tsp(\bm{c}^0)}{\bm{c}^0 \bm{x}_0}$.
    We construct the metric $\bm{c}^1$ on $V_{n+1}$ adding the node $w$ at distance $0$ form $b$: \
    $c^1_{wb} = 0$, \
    $c^1_{vw} = c^0_{vb}$ \ for all nodes $v \in V_n \setminus b$ \ and \
    $c^1_e = c^0_e$ \ for all edges $e \in E_n$.
    Clearly $\bm{c}^1$ is metric and $\bm{c}^1 \bm{x}_1 = \bm{c}^0 \bm{x}_0$.
    
    We first show that $\tsp(\bm{c}^0) \geq \tsp(\bm{c}^1)$.
    Let $\bm{t}_0$ be a tour on $K_n$ that is optimal for $\tsp(\bm{c}^0)$: \ $\tsp(\bm{c}^0) = \bm{c}^0 \bm{t}_0$.
    Let $\bm{t}^+_0$ be the tour on $K_{n+1}$ which retraces the steps of $\bm{t}_0$ but visits $w$ immediately after $b$, that is, if the sequence of nodes visited by $\bm{t}_0$ is \ $b, v_2, \dots, v_n$, \ then the sequence of nodes visited by $\bm{t}^+_0$ is \ $b, w, v_2, \dots, v_n$. \
    It is immediate to verify that $\bm{c}^1 \bm{t}^+_0 = \bm{c}^0 \bm{t}_0$: \
    $\bm{t}^+_0$ uses the same edges of $\bm{t}_0$, except for $aw$ and $wb$ in place of $ab$, which still involve the same cost $c^1_{aw} + c^1_{wb} = c^0_{ab} + 0$. 
    Therefore $\tsp(\bm{c}^1) \leq \bm{c}^1 \bm{t}^+_0 = \bm{c}^0 \bm{t}_0 = \tsp(\bm{c}^0)$.
    
    On the other hand, $\tsp(\bm{c}^1)$ is always larger than $\tsp(\bm{c}^0)$: given an optimal tour $\bm{t}_1$ for $\tsp(\bm{c}^1)$, we can always cut out $w$ to get a tour $\bm{t}^-_1$ with $\bm{c}^0 \bm{t}^-_1 \leq \bm{c}^1 \bm{t}_1$ by triangle inequality, thus \ $\tsp(\bm{c}^0) \leq \bm{c}^0 \bm{t}^-_1 \leq \bm{c}^1 \bm{t}_1 = \tsp(\bm{c}^1)$.
    
    Putting both the inequalities together, we have $\tsp(\bm{c}^1) = \tsp(\bm{c}^0)$, which finally proves
    $\gapp(\bm{x}_1) \geq \frac{\tsp(\bm{c}^1)}{\bm{c}^1 \bm{x}_1} = \frac{\tsp(\bm{c}^0)}{\bm{c}^0 \bm{x}_0} = \gapp(\bm{x}_0)$.
    \qed
\end{proof}

\begin{proof}[of \Cref{lem:bound_ancestors}]
    Let $\bm{x}'$ be a successor of $\bm{x}_0$.
    Let $e_1, \dots, e_p$ be the $1$-edges of $\bm{x}_0$ \ and let $d_1, \dots, d_p$ and $d'_1, \dots, d'_p$ be the lengths of the corresponding $1$-paths of $\bm{x}$ and $\bm{x}'$, respectively.
    Consider the successor $\tilde{\bm{x}}$ of $\bm{x}_0$ whose $1$-paths have $\max(d_1, d'_1), \dots, \max(d_p, d'_p)$ internal nodes.
    Then $\tilde{\bm{x}}$ is a successor of $\bm{x}'$, hence \ $\gapp(\bm{x}') \leq \gapp(\tilde{\bm{x}})$ \ for \Cref{lem:bb_move_gap}, \
    and $\tilde{\bm{x}}$ is a successor of $\bm{x}$ as well, hence \ $\gapp(\tilde{\bm{x}}) \leq \gb(\bm{x})$ \ for \Cref{thm:GB_algo}.
    This proves the lemma.
    \qed
\end{proof}

%%%%%%%%%% Proof of Lemma
\subsection{Proof of \texorpdfstring{\Cref{lem:doptp_ii}}{Lemma}} \label{sec:proof_opti_2}

In Section \ref{sec:doptp_ii}, we introduced two different LPs:
$\dual\optp$ \eqref{eq:doptp_obj} -- \eqref{eq:doptp_var_mu} \
and $\dual\optii$ \eqref{eq:doptii_obj} -- \eqref{eq:doptii_var_mu};
in \Cref{lem:doptp_ii}, we claimed that they have the same objective value.
This whole section is devoted to proving this equivalence.

We begin by giving a non-formal interpretation of the equivalence, as it may guide the understanding of the subsequent arguments.
As it appears in the definition of the objective function~\eqref{eq:doptp_obj}, we shall think that the ``important'' variables of $\dual\optp$ are the $\bm{\mu}$, while the $\bm{\lambda}$ are just ``auxiliary''. 
An optimal dual assignment is an assignment of positive weights $\bm{\mu}$ to the tours, such that, for every edge $ij$, the sum of the weights of all the tours passing through $ij$ gives exactly $x_{ij}$ (because of complementary slackness).
If we only considered the (positive) weights $\bm{\mu}$, we would obtain the undesired necessary condition that no tour shall use edges outside the support graph (where $x_{ij} = 0$); $\bm{\lambda}$ variables are meant to ``correct'' this restriction.
In the constraints~\eqref{eq:doptp_constr}, the variable $\lambda_{ijk}$ is subtracted from the edge $ij$ and added to $ik, jk$, as if it is ``erasing'' the passage of a tour on $ij$ and diverting it to pass through $ik$ and $jk$ instead.
In the flavor of this interpretation, the new formulation $\dual\optii$ has $\bm{\mu}$ variables for walks which stay on the support graph and may traverse the same edge multiple times; $\bm{\lambda}$ variables are no longer needed, as their corrections are already taken into account along walks.

In the subsequent proofs, we see how to pass from one formulation to the other, transforming the $\bm{\mu}$ variables of $\dual\optp(\bm{x})$, indexed by tours on $K_n$, into $\bm{\mu}$ variables of $\dual\optii(\bm{x})$, indexed by walks on $G_{\bm{x}}$, and vice-versa. \
To convert a tour into a walk on $G_{\bm{x}}$, we substitute edges $ij$ not in $G_{\bm{x}}$ with paths connecting $i$ and $j$ in $G_{\bm{x}}$%
\footnote{This is always possible since $G_{\bm{x}}$ is connected, for subtour elimination constraints.}.
To convert a walk into a tour, we consider the sequence of the nodes visited along the walk (having fixed an order and a starting point) and cut out repeated visits to the same nodes.

To accomplish the proof of \Cref{lem:doptp_ii}, we introduce an intermediate formulation, \ $\dual\opti$.
For a given vertex $\bm{x}$, $\dual\opti(\bm{x})$ is defined as:
{\small
\begin{align}
    \text{maximize \ \ }
        & \sum_{\bm{w} \text{ walk}} \mu_{\bm{w}} \\
    \text{subject to: \ }
        & \sum_{k \neq i, j} \! (-\! \lambda_{ijk} +\! \lambda_{ikj} +\! \lambda_{jki})
            + \!\! \sum_{\bm{w} \text{ walk}} \!\! w_{ij} \mu_{\bm{w}}
            \ \leq x_{ij}
            & & \forall ij \in E_n,
            \label{eq:dopti_constr} \\
        & \lambda_{ijk} \ \geq 0
            & & \forall ij \in E_n, \ k \neq i, j, \\
        & \mu_{\bm{w}} \ \geq 0
            & & \forall \bm{w} \text{ walk on $K_n$}.
\end{align}}

Since both tours on $K_n$ and walks on $G_{\bm{x}}$ are special cases of walks on $K_n$, $\dual\opti(\bm{x})$ is an extended version of both $\dual\optp$ and $\dual\optii(\bm{x})$.

\begin{remark}
    Before proving the equivalence of $\dual\optp$, $\dual\opti$, and $\dual\optii(\bm{x})$ (that is, they give the same optimal value), let us address an ``exception'' we may encounter along the proofs.
    In the following arguments, we often construct new walks by modifying pre-existing ones, adding or removing edges.
    In principle, it is not excluded that a new walk $\bm{w}'$ obtained this way passes through an edge $ij$ more than twice.
    The problem arises because, as stated in~\Cref{rem:walk_mult}, we are not considering such walks in our formulations.
    Whenever this situation occurs, we shall then replace $\bm{w}'$ with another walk $\bm{w}''$ obtained removing two copies of $ij$; in addition, if a variable $\mu_{\bm{w}'}$ is assigned to $\bm{w}'$ in the $\dual\opti$ problem, we shall instead add the same amount to the variable $\mu_{\bm{w}''}$ and not consider $\mu_{\bm{w}'}$ at all.
    It is crucial to notice that this change preserves the same objective value as the ``erroneous'' construction.
    It preserves the feasibility of the assignment as well, since the value of the constraints~\eqref{eq:dopti_constr} stays the same on all the edges but $ij$, on which it eventually decreases (since we dropped two copies of $ij$ in $\bm{w}''$).
\end{remark}

We are now ready to prove the two main lemmas of this section.

\begin{lemma} \label{lem:dopti_p}
    $\dual\opti(\bm{x}) = \dual\optp(\bm{x})$ \ for every vertex $\bm{x}$ of $\psep{n}$.
\end{lemma}

\begin{proof}
    Since $\dual\optp(\bm{x})$ is a special case of $\dual\opti(\bm{x})$, we immediately have \ $\dual\opti(\bm{x}) \geq \dual\optp(\bm{x})$ \ and we only need to prove \ $\dual\opti(\bm{x}) \leq \dual\optp(\bm{x})$.
    We prove it by showing that, for any given optimal solution of $\dual\opti(\bm{x})$, we can compute a feasible solution of $\dual\optp(\bm{x})$ with the same objective value.
    
    Let $(\bm{\lambda}^0, \bm{\mu}^0)$ be an optimal solution for $\dual\opti(\bm{x})$.
    Let $\bm{w}^0$ be a walk on $K_n$.
    If $\bm{w}^0$ is also a tour, then we can leave the variable $\mu_{\bm{w}^0}$ as it is;
    otherwise, if $\bm{w}^0$ is not a tour, in the aim of designing an assignment for $\dual\optp(\bm{x})$, we have to set $\mu_{\bm{w}^0}$ to $0$. 
    To do so without reducing the optimal value, we also need to adjust a couple of other variables.
    Let $c$ be a node that is encountered more than once along $\bm{w}^0$ (it exists since $\bm{w}^0$ is not a tour), and let $a$ and $b$ be the nodes encountered immediately before and after $c$ the second time it is traversed.
    Consider the walk $\bm{w}^1$ that retraces the steps of $\bm{w}^0$ but walks through the shortcut $a, b$ instead of $a, c, b$, as illustrated in \Cref{fig:dopt_lemmas}.
    We design a new assignment of variables $(\bm{\lambda}^1, \bm{\mu}^1)$, introducing, for simplicity, the constant $\delta^0 := \mu^0_{\bm{w}^0}$:
    $$
    \begin{array}{l@{\ := \ }ll}
        \lambda^1_{abc} & \lambda^0_{abc} + \delta^0, \\[3pt]
        \lambda^1_{ijk} & \lambda^0_{ijk}
            & \forall ijk \neq abc ,
    \end{array}
    \qquad
    \begin{array}{l@{\ := \ }ll}
        \mu^1_{\bm{w}^0} & \mu^0_{\bm{w}^0} - \delta^0 = 0, \\[3pt]
        \mu^1_{\bm{w}^1} & \mu^0_{\bm{w}^1} + \delta^0, \\[3pt]
        \mu^1_{\bm{w}}  & \mu^0_{\bm{w}}
            & \forall \bm{w} \neq \bm{w}^0, \bm{w}^1.
    \end{array}
    $$

    It is immediate to check that the objective value does not change:
    {\small
    $$
    \everymath{\displaystyle}
    \begin{array}{ll}
        \sum_{\bm{w}} \mu^1_{\bm{w}}
            & = \ \bigg( \sum_{\bm{w} \neq \bm{w}^0, \bm{w}^1} \!\! \mu^1_{\bm{w}} \bigg)      + \mu^1_{\bm{w}^0} + \mu^1_{\bm{w}^1} \\
            & = \ \bigg( \sum_{\bm{w} \neq \bm{w}^0, \bm{w}^1} \!\! \mu^0_{\bm{w}} \bigg) + 0 + (\mu^0_{\bm{w}^1} \!+\! \mu^0_{\bm{w}^0})
            \ = \ \sum_{\bm{w}} \mu^0_{\bm{w}} \ .
    \end{array}
    $$}

    We now show that $(\bm{\lambda}^1, \bm{\mu}^1)$ is feasible for $\dual\opti$, that is, all the constraints~\eqref{eq:dopti_constr} are satisfied.
    In fact, the values of the constraints attained by $(\bm{\lambda}^1, \bm{\mu}^1)$ is exactly the same as the ones attained by $(\bm{\lambda}^0, \bm{\mu}^0)$.
    We check this fact for the constraint associated with the edge $ab$;
    the equalities holds simply by definition of $(\bm{\lambda}^1, \bm{\mu}^1)$ and considering that the multiplicities of the walk $\bm{w}^1$ are given as \
    $w^1_{ab} = w^0_{ab} \!+\! 1, \ \
    w^1_{ac} = w^0_{ac} \!-\! 1, \ \
    w^1_{cb} = w^0_{cb} \!-\! 1$.

    {\small
    $$
    \everymath{\displaystyle}
    \begin{array}{l@{\qquad}cl}
        \multicolumn{3}{l}{\sum_{k \neq a, b}
                (-\lambda^1_{abk} +\lambda^1_{akb} +\lambda^1_{bka})
                \ + \ \sum_{\bm{w}} w_{ab} \mu^1_{\bm{w}} = } \\
            
            & = & \bigg( \sum_{k \neq a, b, c} \!\!
                (-\lambda^1_{abk} +\lambda^1_{akb} +\lambda^1_{bka}) \bigg)
                + (-\lambda^1_{abc} +\lambda^1_{acb} +\lambda^1_{cba}) \\
            &   & + \bigg( \sum_{\bm{w} \neq \bm{w}^0, \bm{w}^1} \!\!
                w_{ab} \mu^1_{\bm{w}} \bigg)
                + w^0_{ab} \mu^1_{\bm{w}^0} + w^1_{ab} \mu^1_{\bm{w}^1} \\
            
            & = & \bigg( \sum_{k \neq a, b, c} \!\!
                (-\lambda^0_{abk} +\lambda^0_{akb} +\lambda^0_{bka}) \bigg)
                + ( -(\lambda^0_{abc} \!+\! \delta^0) +\lambda^0_{acb} +\lambda^0_{cba} ) \\
            &   & + \bigg( \sum_{\bm{w} \neq \bm{w}^0, \bm{w}^1} \!\!
                w_{ab} \mu^0_{\bm{w}} \bigg)
                + 0 + w^1_{ab} (\mu^0_{\bm{w}^1} + \delta^0) \\
            
            & = & \bigg( \sum_{k \neq a, b} \!\!
                (-\lambda^0_{abk} +\lambda^0_{akb} +\lambda^0_{bka}) \bigg)
                \ - \ \delta^0 \\
            &   & + \bigg( \sum_{\bm{w} \neq \bm{w}^0, \bm{w}^1} \!\!
                w_{ab} \mu^0_{\bm{w}} \bigg)
                + w^1_{ab} \mu^0_{\bm{w}^1} + (w^0_{ab} \!+\! 1) \delta^0 \\
            
            & = & \bigg( \sum_{k \neq a, b} \!\!
                (-\lambda^0_{abk} +\lambda^0_{akb} +\lambda^0_{bka}) \bigg)
                \ - \ \delta^0 \\
            &   & + \bigg( \sum_{\bm{w} \neq \bm{w}^0, \bm{w}^1} \!\!
                w_{ab} \mu^0_{\bm{w}} \bigg)
                + w^1_{ab} \mu^0_{\bm{w}^1} + w^0_{ab} \mu^0_{\bm{w}^1} + \delta^0 \\
            
            & = & \sum_{k \neq a, b}
                (-\lambda^0_{abk} +\lambda^0_{akb} +\lambda^0_{bka})
                \ + \ \sum_{\bm{w}} w_{ab} \mu^0_{\bm{w}}
    \end{array}
    $$}
    
    For $ac, bc$ and all other edges $ij$, the computation is similar: contributes brought by the new $\bm{\lambda}^1$ and $\bm{\mu}^1$ (i.e. $\pm \delta^0$) cancel out and give the same value of the constraint computed for $(\bm{\lambda}^0, \bm{\mu}^0)$, as illustrated in \Cref{fig:dopt_lemmas}.
    
    \begin{figure}[htbp]
        \centering
        \newcommand{\preparetrianglecoordinates}{
    \node at (-2, -1)   (a) {};
    \node at (+2, -1)   (b) {};
    \node at (0,  +1)   (c) {};
    \node at (+4, -1.5) (d) {};
    \node at (+5, +0)   (e) {};
    \node at (+4, +1.5) (f) {};
    \node at (+2, +1.5) (g) {};
    \node at (-2, +1.5) (h) {};
    \node at (-4, +1.5) (i) {};
    \node at (-5, +0)   (l) {};
    \node at (-4, -1.5) (m) {};
}

\begin{tikzpicture}[x=0.5cm, y=0.5cm]
    \begin{scope}[yshift=+2cm]
        \preparetrianglecoordinates
        \foreach \vname in {a, b, c}{
            \fill (\vname) circle (2pt);
        }
        \node[below] at (a) {$a$};
        \node[below] at (b) {$b$};
        \node[above] at (c) {$c$};
    
        \node at (1pt, 0) (xx) {};
        \node at (0, 1pt) (yy) {};
    
        \foreach \vname in {a, b, c, d, f, g, h, i, m}{
            \node at ($(\vname) + (yy)$) (\vname u) {};
            \node at ($(\vname) - (yy)$) (\vname d) {};
        }
        \foreach \vname in {e, l}{
            \node at ($(\vname) + (xx)$) (\vname r) {};
            \node at ($(\vname) - (xx)$) (\vname l) {};
        }
        
        \draw[blue] plot coordinates {(lr) (mu) (au) (cd) (bu) (du) (el) 
                                           (fd) (gd) (cd) (hd) (id) (lr)};
        \draw[cyan] plot coordinates {(ll) (md) (ad)      (bd) (dd) (er)
                                           (fu) (gu) (cu) (hu) (iu) (ll)};
        
        \node at (0,  0.2) {$\bm{w}^0$};
        \node at (0, -1.5) {$\bm{w}^1$};
    \end{scope}

    \begin{scope}[xshift=-2.2cm]
        \begin{scope}[yshift=2]
            \preparetrianglecoordinates
            \draw[blue] plot coordinates {($0.5*(m) + 0.5*(a)$) (a) (c) (b) ($0.5*(d) + 0.5*(b)$)};
        \end{scope}
        \begin{scope}[yshift=-2]
            \preparetrianglecoordinates
            \draw[cyan] plot coordinates {($0.5*(m) + 0.5*(a)$) (a) (b) ($0.5*(d) + 0.5*(b)$)};
        \end{scope}
        \begin{scope}[scale=0.9]
            \preparetrianglecoordinates
            \draw[orange]         plot coordinates {(a) (c) (b)};
            \draw[orange, dashed] plot coordinates {(a) (b)};
        \end{scope}
        
        \preparetrianglecoordinates
        \foreach \vname in {a, b, c}{
            \fill (\vname) circle (2pt);
        }
        \node[below] at (a) {$a$};
        \node[below] at (b) {$b$};
        \node[above] at (c) {$c$};
        \node        at (-2.8, 1)   {$(\bm{\lambda}^0, \bm{\mu}^0)$};
        \node[right] at (0.7, +0.7) {\small $\mu^0_{\bm{w}_0}$};
        \node[below] at (0, -1.1)   {\small $\mu^0_{\bm{w}_1}$};
        \node[above] at (0, -0.9)   {\small $\lambda^0_{abc}$};
    \end{scope}

    \begin{scope}[xshift=+2.2cm]
        \begin{scope}[yshift=-2]
            \preparetrianglecoordinates
            \draw[cyan] plot coordinates {($0.5*(m) + 0.5*(a)$) (a) (b) ($0.5*(d) + 0.5*(b)$)};
        \end{scope}
        \begin{scope}[scale=0.9]
            \preparetrianglecoordinates
            \draw[orange]         plot coordinates {(a) (c) (b)};
            \draw[orange, dashed] plot coordinates {(a) (b)};
        \end{scope}
        
        \preparetrianglecoordinates
        \foreach \vname in {a, b, c}{
            \fill (\vname) circle (2pt);
        }
        \node[below] at (a) {$a$};
        \node[below] at (b) {$b$};
        \node[above] at (c) {$c$};
        \node        at (-2.8, 1)   {$(\bm{\lambda}^1, \bm{\mu}^1)$};
        \node[right] at (0.7, +0.7) {\small $\mu^1_{\bm{w}_0}$};
        \node[below] at (0, -1.1)   {\small $\mu^1_{\bm{w}_1}$};
        \node[above] at (0, -0.9)   {\small $\lambda^1_{abc}$};
        
        \node[right] at (+2.7, +0.8) {{\scriptsize $\mu^1_{\bm{w}_0} := \mu^0_{\bm{w}_0} \!- \delta^0 = 0$}};
        \node[right] at (+2.7, 0)    {{\scriptsize $\mu^1_{\bm{w}_1} := \mu^0_{\bm{w}_1} \!+ \delta^0$}};
        \node[right] at (+2.7, -0.8) {{\scriptsize $\lambda^1_{abc} := \lambda^0_{abc} \!+ \delta^0$}};
    \end{scope}
\end{tikzpicture}
        \caption{Graphical representation of the proof of \Cref{lem:dopti_p}.
        Above: the walks $\bm{w}^0$ and $\bm{w}^1$: the latter is obtained from the former by cutting $c$ out of the path $acb$.
        Below: a detail of $\bm{\lambda}$ and $\bm{\mu}$ variables, represented as triangles and walks with solid or dashed edges according to whether the variable is added or subtracted in the corresponding constraint.
        For both the assignments $(\bm{\lambda}^0, \bm{\mu}^0)$ and $(\bm{\lambda}^1, \bm{\mu}^1)$, the constraint~\eqref{eq:dopti_constr} attains the same value on every edge.}
        \label{fig:dopt_lemmas}
    \end{figure}

    We apply the same argument on $\bm{w}^1$ and go on iteratively, considering the walks $\bm{w}^1, \dots, \bm{w}^s$, until every node is encountered just once along the last walk.
    Eventually, $\bm{w}^s$ is a tour and all the variables $\mu_{\bm{w}^0}, \mu_{\bm{w}^1}, \dots, \mu_{\bm{w}^{s-1}}$ are $0$.

    Applying this whole procedure for all the walks that are not tours yields a solution $(\bm{\mu}^*, \bm{\lambda}^*)$ for $\dual\opti(\bm{x})$ which attains the optimal value, and with $\mu^*_{\bm{w}} = 0$ if $\bm{w}$ is not a tour; such a solution is in turn a feasible solution for $\dual\optp(\bm{x})$.
    This completes the proof.
    \qed
\end{proof}

\begin{lemma} \label{lem:dopti_ii}
    $\dual\opti(\bm{x}) = \dual\optii(\bm{x})$ \ for every vertex $\bm{x}$ of $\psep{n}$.
\end{lemma}

\begin{proof}
    Since $\dual\optii(\bm{x})$ is a special case of $\dual\opti(\bm{x})$,
    we immediately have \ $\dual\opti(\bm{x}) \geq \dual\optii(\bm{x})$ \ and we only need to prove \ $\dual\opti(\bm{x}) \leq \dual\optii(\bm{x})$.
    We prove it by showing that, for any given optimal solution of $\dual\opti(\bm{x})$, we can compute a feasible solution of $\dual\optii(\bm{x})$ with the same objective value.
    
    Let $(\bm{\lambda}^0, \bm{\mu}^0)$ be an optimal solution for $\dual\opti(\bm{x})$.
    Consider the polytope
    $$
    \begin{array}{rl}
        P :=
            & \left\{ (\bm{\lambda}, \bm{\mu}) \ \mid \
                \begin{array}{l}
                    (\bm{\lambda}, \bm{\mu}) \text{ optimal solution for } \dual\opti(\bm{x}), \\
                    \sum_{ijk} \lambda_{ijk} \leq \sum_{ijk} \lambda^0_{ijk}
                \end{array} \ \right\} \\[12pt]

        =   & \left\{ (\bm{\lambda}, \bm{\mu}) \ \mid \
                \begin{array}{l}
                    (\bm{\lambda}, \bm{\mu}) \text{ feasible solution for } \dual\opti(\bm{x}), \\
                    \sum_{\bm{w}} \mu_{\bm{w}} = \sum_{\bm{w}} \mu^0_{\bm{w}}, \\
                    \sum_{ijk} \lambda_{ijk} \leq \sum_{ijk} \lambda^0_{ijk}
                \end{array} \ \right\} \ .
    \end{array}
    $$
    $P$ is a bounded polyhedron, hence it is compact.
    Consider the function $f: P \rightarrow \R$ which maps $(\bm{\lambda}, \bm{\mu})$ into $f(\bm{\lambda}, \bm{\mu}) := \sum_{ijk} \lambda_{ijk}$.
    Since $f$ is continuous on $P$ compact, by the Weierstrass theorem, there exists at least one point $(\bm{\lambda}^*, \bm{\mu}^*) \in P$ where $f$ attains its minimum value, i.e.,
    $
    f(\bm{\lambda}^*, \bm{\mu}^*) = \min_{(\bm{\lambda}, \bm{\mu}) \in P} f(\bm{\lambda}, \bm{\mu})
    $.
    The following claim, whose proof is deferred at the end of the main argumentation, guarantees that the minimum of $f$ is realized with $\bm{\lambda}^* = \bm{0}$.
    \begin{claim}
        For any $(\bm{\lambda}, \bm{\mu})$ feasible solution for $\dual\opti(\bm{x})$ with $\bm{\lambda} \neq \bm{0}$, there exists another feasible solution $(\bar{\bm{\lambda}}, \bar{\bm{\mu}})$ for $\dual\opti(\bm{x})$ such that \ $\sum_{\bm{w}} \bar{\mu}_{\bm{w}} = \sum_{\bm{w}} \mu_{\bm{w}}$ \ and \ $\sum_{ijk} \bar{\lambda}_{ijk} < \sum_{ijk} \lambda_{ijk}$.
    \end{claim}
    Moreover, for any edge $ij$ not in $E_{\bm{x}}$, \ $x_{ij} = 0$ and the constraint~\eqref{eq:dopti_constr} becomes $\sum_{w} w_{ij} \mu^*_{\bm{w}} \leq 0$, thus $\mu^*_{\bm{w}} = 0$ for every walk $\bm{w}$ with $w_{ij} > 0$.
    This ultimately means that $\lambda^*_{ijk} = 0$ for all $ijk$ and $\mu^*_{\bm{w}} = 0$ for all the walks $\bm{w}$ which do not stay on the support graph $G_{\bm{x}}$; that is, the optimal solution $(\bm{\lambda}^*, \bm{\mu}^*)$ of $\dual\opti(\bm{x})$ is in turn a feasible solution of $\dual\optii(\bm{x})$ with the same objective value.
    \qed
\end{proof}

\begin{proof}[of the claim]
    We divide the argument into two cases.

    Case 1: exists an edge $ab$ such that exist a node $c$ with $\lambda_{abc} > 0$ and a walk $\bm{w}'$ with $w'_{ab} \mu_{\bm{w}'} > 0$ (the walk traverses $ab$ at least once and the associated variable $\mu_{\bm{w}}$ is not zero).
    We aim to adjust the assignment by a small quantity $\delta$ that leaves $\lambda_{abc}$ and $\mu_{\bm{w}'}$ grater or equal to $0$: \
    $\delta := \min(\lambda_{abc}, \ \mu_{\bm{w}'})$.
    Let $\bar{\bm{w}}$ be the walk that retraces the steps of $\bm{w}'$ but drops once the edge $ab$ and walks through $a, c, b$ instead.
    We design a new assignment of variables:
    $$
    \begin{array}{l@{\ := \ }l@{\quad}l}
        \bar{\lambda}_{abc} & \lambda_{abc} - \delta \ , \\[3pt]
        \bar{\lambda}_{ijk} & \lambda_{ijk}
            & \forall ijk \neq abc \ ,
    \end{array}
    \qquad \qquad \qquad
    \begin{array}{l@{\ := \ }l@{\quad}l}
        \bar{\mu}_{\bm{w}'}      & \mu_{\bm{w}'} - \delta \ , \\[3pt]
        \bar{\mu}_{\bar{\bm{w}}} & \mu_{\bar{\bm{w}}} + \delta \ , \\[3pt]
        \bar{\mu}_{\bm{w}}       & \mu_{\bm{w}}
            & \forall \bm{w} \neq \bm{w}', \bar{\bm{w}} \ .
    \end{array}
    $$
    It is immediate to check that the objective value stays the same and the sum of the $\bm{\lambda}$ variables strictly decreases:
    $$
    \everymath{\displaystyle}
    \begin{array}{c}
        \sum_{\bm{w}} \bar{\mu}_{\bm{w}}
            = \sum_{\bm{w} \neq \bm{w}', \bar{\bm{w}}} \!\!\! \bar{\mu}_{\bm{w}} \
            + \bar{\mu}_{\bm{w}'} + \bar{\mu}_{\bar{\bm{w}}}
            = \sum_{\bm{w} \neq \bm{w}', \bar{\bm{w}}} \!\!\! \mu_{\bm{w}} \
            + \mu_{\bm{w}'} \!-\! \delta + \mu_{\bar{\bm{w}}} \!+\! \delta
            = \sum_{\bm{w}} \mu_{\bm{w}} \ , \\[5pt]
        \sum_{ijk} \bar{\lambda}_{ijk}
            = \sum_{ijk \neq abc} \!\!\! \bar{\lambda}_{ijk} \
            + \bar{\lambda}_{abc}
            = \sum_{ijk \neq abc} \!\!\! \lambda_{ijk} \
            + \lambda_{abc} \!-\! \delta
            = \sum_{ijk} \lambda_{ijk} - \delta \ .
    \end{array}
    $$
    Showing that the new assignment $(\bar{\bm{\lambda}}, \bar{\bm{w}})$ is feasible for $\dual\opti(\bm{x})$, as in the proof of \Cref{lem:dopti_p}, is just a matter of expand the new variables and check that the new weights and multiplicities balance out to give the same value as $(\bm{\lambda}, \bm{\mu})$ for all the constraints.

    Case 2: for every edges $ij$, all the variables $\lambda_{ijk}$ are $0$ \ or all the walks $\bm{w}$ have $w_{ij} \mu_{\bm{w}} = 0$.
    Then, for every edge $ij$, \ either \
    $\sum_{\bm{w}} w_{ij} \mu_{\bm{w}}
    \leq \sum_{\bm{w}} w_{ij} \mu_{\bm{w}} + \sum_{k \neq i, j}(-0 +\lambda_{ikj} +\lambda_{jki})
    = \sum_{\bm{w}} w_{ij} \mu_{\bm{w}} + \sum_{k \neq i, j}(-\lambda_{ijk} +\lambda_{ikj} +\lambda_{jki})
    \leq x_{ij}$, \
    or \ $\sum_{\bm{w}} w_{ij} \mu_{\bm{w}} = 0 \leq x_{ij}$.
    This proves that we can remove all the $\bm{\lambda}$ variables (i.e., set them to $0$) to obtain the new assignment $(\bm{0}, \bm{\mu})$, which is feasible and preserves all the $\bm{\mu}$ variables, thus preserving the objective value.

    Notice that this completes the proof since the two cases analyzed are the logical negation of each other.
    Writing the two clauses with formal predicate logic, it is immediate to check that \
    $\lnot \ (\text{Case 1}) \ \Leftrightarrow \ (\text{Case 2})$, as we have:
    $$
    \begin{array}{lrl}
        \text{Case 1: \ }
            & \exists ab \in E_n\!
            & : ( \
                \exists c \in V_n : \lambda_{abc} > 0 \ \land \
                \exists \bm{w}' \text{ walk on } K_n : w'_{ab} \mu_{\bm{w}'} > 0 \ ) \ , \\[3pt]
        \text{Case 2: \ }
            & \forall ij \in E_n\!
            & : ( \
                \forall k \in V_n : \lambda_{ijk} = 0 \ \lor \
                \forall \bm{w} \text{ walk on } K_n : w_{ij} \mu_{\bm{w}} = 0 \ ) \ .
                \ \ \qed
    \end{array}
    $$
\end{proof}

Finally, \Cref{lem:doptp_ii} trivially follows, concatenating \Cref{lem:dopti_p} and \Cref{lem:dopti_ii}.

%%%%%%%%%% Proof of the soundness of the GB algorithm
\subsection{Proof of the soundness of the \texorpdfstring{\gb}{GB} algorithm} \label{subsec:tech_GB}

In \Cref{subsec:bounding_gap}, we introduced the \gb\ algorithm and proved its soundness relying upon claims that we only stated.
In this section, we provide proofs of all the technical details, ultimately completing the demonstration of \Cref{thm:GB_algo}.

Recall how we designed the \gb\ algorithm.
Starting from a vertex $\bm{x}_0$ alongside with an optimal solution $\bm{\mu}^0$ for the dual problem $\dual\optii(\bm{x}_0)$, we considered the successor $\bm{x}_d$ obtained expanding a $1$-edge $ab$  of $\bm{x}_0$ to a $1$-path with $d$ internal nodes and we constructed an assignment $\bm{\mu}^d$ for $\dual\optii(\bm{x}_d)$.
To do so, for every walk $\bm{w}^0$ on $G_{\bm{x}_0}$, depending on its multiplicity on the edge $ab$, we considered different walks on $G_{\bm{x}_d}$: \
for $\bm{w}^0 \in \cW^{ab}_0$, \ we considered $d \!+\! 1$ walks $\bm{w}^d_k$ on $G_{\bm{x}_d}$ and set \ $\mu^d_{\bm{w}^d_k} := \frac{1}{d+1} \mu^0_{\bm{w}^0}$ \ for all $k = 0, \dots, d$; \
for $\bm{w}^0 \in \cW^{ab}_1$ \ or \ $\bm{w}^0 \in \cW^{ab}_2$, \ we considered one walk $\bm{w}^d$ on $G_{\bm{x}_d}$ and set \ $\mu^d_{\bm{w}^d} := \mu^0_{\bm{w}^0}$ (see \Cref{subsec:bounding_gap}).

\begin{claim}
    (i) $\bm{\mu}^d$ attains the same objective value (in $\dual\optii(\bm{x}_d)$) \ as $\bm{\mu}^0$ (in $\dual\optii(\bm{x}_0)$); \
    (ii) $\bm{\mu}^d$ satisfies all the constraints~\eqref{eq:doptii_constr}, except those associated to the edges of the $1$-path $a a_1 \dots a_d b$, which nonetheless remain bounded by a fixed constant.
\end{claim}

\begin{proof}
    The proof of the first fact (i) is the following straightforward computation:
    {\small
    $$
    \everymath{\displaystyle}
    \begin{array}{rl}
        \sum_{\bm{w}^d \text{ walk on } G_{\bm{x}_d}} \mu^d_{\bm{w}^d}
             & = \sum_{\bm{w}^0 \in \cW^{ab}_0} \Big( \sum_{k=0}^d \mu^d_{\bm{w}^d_k} \Big)
                + \sum_{\bm{w}^0 \in \cW^{ab}_1} \mu^d_{\bm{w}^d}
                + \sum_{\bm{w}^0 \in \cW^{ab}_2} \mu^d_{\bm{w}^d} \\[18pt]
             & = \sum_{\bm{w}^0 \in \cW^{ab}_0} (d \!+\! 1) \frac{1}{d \!+\! 1} \mu^0_{\bm{w}^0}
                + \sum_{\bm{w}^0 \in \cW^{ab}_1} \mu^0_{\bm{w}^0}
                + \sum_{\bm{w}^0 \in \cW^{ab}_2} \mu^0_{\bm{w}^0} \\[18pt]
            & = \sum_{\bm{w}^0 \text{ walk on } G_{\bm{x}_0}} \mu^0_{\bm{w}^0} \ .
    \end{array}
    $$}

    To prove the second fact (ii), we consider the edges of $E_{\bm{x}_d}$ and study the value of the associated constraints~\eqref{eq:doptii_constr}.

    For edges $ij$ of $E_{\bm{x}_d}$ not in the $1$-path $a a_1 \dots b$, the multiplicity of the new walks $\bm{w}^d$ are the same of the ones they were originated from, i.e., \ $w^d_{ij} = w^0_{ij}$, \ and the constraint is still satisfied:
    {\small
    $$
    \everymath{\displaystyle}
    \begin{array}{rl}
        \sum_{\bm{w}^d \text{ walk on } G_{\bm{x}_d}} \!\!\!\!\!\!\!\!\!\!
            w^d_{ij} \, \mu^d_{\bm{w}^d}
             & = \sum_{\bm{w}^0 \in \cW^{ab}_0} \!\!\!\!
                \Big( \sum_{k=0}^d (w^d_k)_{ij} \, \mu^d_{\bm{w}^d_k} \Big)
                + \sum_{\bm{w}^0 \in \cW^{ab}_1} \!\!\!\!\! w^d_{ij} \, \mu^d_{\bm{w}^d}
                + \sum_{\bm{w}^0 \in \cW^{ab}_2} \!\!\!\!\! w^d_{ij} \, \mu^d_{\bm{w}^d} \\[18pt]
             & = \sum_{\bm{w}^0 \in \cW^{ab}_0} \!\!\!\!
                \Big( (d \!+\! 1) \ (w^0_{ij} \ \frac{1}{d \!+\! 1} \mu^0_{\bm{w}^0}) \Big)
                + \sum_{\bm{w}^0 \in \cW^{ab}_1} \!\!\!\!\! w^0_{ij} \, \mu^0_{\bm{w}^0}
                + \sum_{\bm{w}^0 \in \cW^{ab}_2} \!\!\!\!\! w^0_{ij} \, \mu^0_{\bm{w}^0} \\[18pt]
             & = \sum_{\bm{w}^0 \text{ walk on } G_{\bm{x}_0}} \!\!\!\!\!\!\!\!\!\!
                w^0_{ij} \, \mu^0_{\bm{w}^0}
                \ \leq \ (x_0)_{ij} \ = \ (x_d)_{ij} \ .
    \end{array}
    $$}

    It only remains to study the edges of the $1$-path and bound from above the value of the constraints associated with them. 
    Consider, for instance, $a a_1$ (analogous computations apply to the other edges):
    {\small
    $$
    \everymath{\displaystyle}
    \begin{array}{rl}
        \sum_{\bm{w}^d \text{ walk on $G_{\bm{x}_d}$}} \!\!\!\!\!\!\!\!\!\!
            w^d_{a a_1} \, \mu^d_{\bm{w}^d}
             & = \sum_{\bm{w}^0 \in \cW^{ab}_0} \!\!\!\!
                \Big(\sum_{k=0}^d (w^d_k)_{a a_1} \, \mu^d_{\bm{w}^d_k} \Big)
                + \sum_{\bm{w}^0 \in \cW^{ab}_1} \!\!\!\!\! w^d_{a a_1} \, \mu^d_{\bm{w}^d}
                + \sum_{\bm{w}^0 \in \cW^{ab}_2} \!\!\!\!\! w^d_{a a_1} \, \mu^d_{\bm{w}^d} \\[18pt]
             & = \sum_{\bm{w}^0 \in \cW^{ab}_0} \!\!\!\!
                \Big( 0 + d \cdot 2 \, \frac{1}{d \!+\! 1} \mu^0_{\bm{w}^0} \Big)
                + \sum_{\bm{w}^0 \in \cW^{ab}_1} \!\!\!\!\! 1 \, \mu^0_{\bm{w}^0}
                + \sum_{\bm{w}^0 \in \cW^{ab}_2} \!\!\!\!\! 2 \, \mu^0_{\bm{w}^0} \\[18pt]
             & \leq \ \ 2 \!\!\! \sum_{\bm{w}^0 \in \cW^{ab}_0} \!\!\! \mu^0_{\bm{w}^0}
                \ + \sum_{\bm{w}^0 \in \cW^{ab}_1} \!\!\! \mu^0_{\bm{w}^0}
                \ + \ 2 \!\!\! \sum_{\bm{w}^0 \in \cW^{ab}_2} \!\!\! \mu^0_{\bm{w}^0} \ .
    \end{array}
    $$}
    The value of the constraint is bounded from above by a quantity that does not depend on $d$.
    \qed
\end{proof}

We denote by $C(\bm{x}_0, ab)$ the right-hand side of the final inequality in the previous proof, which is the constant mentioned in the claim:
$$
C(\bm{x}_0, ab) :=
2 \! \sum_{\bm{w}^0 \in \cW^{ab}_0} \mu^0_{\bm{w}^0} +
     \sum_{\bm{w}^0 \in \cW^{ab}_1} \mu^0_{\bm{w}^0} +
2 \! \sum_{\bm{w}^0 \in \cW^{ab}_2} \mu^0_{\bm{w}^0} \ .
$$
It may be used to rescale $\bm{\mu}^d$ and get the feasible assignment \
$\bm{\mu}^* := \frac{1}{C(\bm{x}_0, ab)} \bm{\mu}^d$ \
for $\dual\optii(\bm{x}_d)$.
Crucially, $C(\bm{x}_0, ab)$ depends only on $\bm{x}_0$: the dependency on $d$ was eliminated in the last inequality by bounding $\frac{d}{d+1}$ by $1$.
Therefore, the constant found is the same for all the vertices obtained by expanding the $1$-edge $ab$ of $\bm{x}_0$ to a $1$-path of arbitrary length.

We now need to extend the construction for a generic successor of $\bm{x}_0$.
To this end, we repeat the previous procedure for all the $1$-edges $e_1, \dots, e_p$ of $\bm{x}_0$, sequentially expanding them to $1$-paths with $d_1, \dots, d_p$ internal nodes.
Let $\bm{x}^1, \dots, \bm{x}^p$ be the vertices obtained along this process.
Accordingly, let $\bm{\mu}^1, \dots, \bm{\mu}^p$ be the assignments of dual variables iteratively originated (starting from $\bm{\mu}^0$).

\begin{claim}
    (i) $\bm{\mu}^p$ attains the same objective value (in $\dual\optii(\bm{x}^p)$) \ as $\bm{\mu}^0$ (in $\dual\optii(\bm{x}_0)$); \
    (ii) $\bm{\mu}^p$ satisfies all the constraints~\eqref{eq:doptii_constr}, except those associated to the $1$-edges, which nonetheless remain bounded by fixed constants.
\end{claim}

\begin{proof}
    It follows directly from the iterative application of the prior claim.
    \qed
\end{proof}

The constants used as bounds for constraints~\eqref{eq:doptii_constr} are computed sequentially at each step $h = 1, \dots, p$ of the construction, starting from the preceding vertex $\bm{x}^{h-1}$ \ (with $\bm{x}^0 := \bm{x}_0$):
$$
C(\bm{x}^{h-1}, e_h) :=
2 \! \sum_{\bm{w}^{h-1} \in \cW^{e_h}_0} \mu^{h-1}_{\bm{w}^{h-1}} +
     \sum_{\bm{w}^{h-1} \in \cW^{e_h}_1} \mu^{h-1}_{\bm{w}^{h-1}} +
2 \! \sum_{\bm{w}^{h-1} \in \cW^{e_h}_2} \mu^{h-1}_{\bm{w}^{h-1}} \ .
$$

To prove the soundness of the \gb\ algorithm, it only remains to show that the factors $C(\bm{x}^{h-1}, e_h)$ are not determined by the order in which the $1$-edges are expanded: they can be computed directly from the starting vertex $\bm{x}^0$.

\begin{claim} \label{clm:C_constants}
    $C(\bm{x}^{h-1}, e_h) = C(\bm{x}^0, e_h)$.
\end{claim}

\begin{proof}
    Before digging into the main argument of the proof, we introduce a notation that prevents ambiguities.
    For $m = 0, 1, 2$, we use \ $\ee^{e_h}_m(\bm{w}^{h-1})$ \ (eventually with subscript $k$ when $m = 0$) to denote the new walks obtained when extending $\bm{w}^{h-1}$ (``\ee'' for ``extend'') to pass through all the new nodes added in the expansion of $e_h$ to a $1$-path with $d_h$ internal nodes (see \Cref{subsec:bounding_gap}).
    With this new notation, the new assignment of $\bm{\mu}$ variables becomes \
    $\mu^h_{\ee^{e_h}_0(\bm{w}^{h-1})_k} := \frac{1}{d_h+1} \mu^{h-1}_{\bm{w}^{h-1}}$ \ for all $k = 0, \dots, d_h$, \ \
    $\mu^h_{\ee^{e_h}_1(\bm{w}^{h-1})} := \mu^{h-1}_{\bm{w}^{h-1}}$, \ or \
    $\mu^h_{\ee^{e_h}_2(\bm{w}^{h-1})} := \mu^{h-1}_{\bm{w}^{h-1}}$, \
    according to whether \ $\bm{w}^{h-1} \in \cW^{e_h}_0$, \ $\bm{w}^{h-1} \in \cW^{e_h}_1$, \ or $\bm{w}^{h-1} \in \cW^{e_h}_2$, respectively.
    
    We also introduce an operator that mimics an \emph{if} statement: we use $\langle condition \rangle$, which has value $1$ if $condition$ is true and $0$ otherwise.
    $C(\bm{x}^{h-1}, e_h)$ can thus be rewritten compactly as
    $$
    C(\bm{x}^{h-1}, e_h) =
    \sum_{\substack{\bm{w}^{h-1} \\ \text{ walk on } G_{\bm{x}^{h-1}}}} \!\!\!
        \sum_{m = 0, 1, 2}
        \theta_m \ \langle \bm{w}^{h-1} \in \cW^{e_h}_m \rangle \ \mu^{h-1}_{\bm{w}^{h-1}} \ ,
    $$
    where $\theta_0 := 2, \ \theta_1 := 1, \ \theta_2 := 2$.    

    To prove the claim, we adopt a recursive strategy: we show that the constant $C(\bm{x}^{h-1}, e_h)$ can be computed by swapping steps $h-1$ and $h$ in the construction, namely \
    $C(\bm{x}^{h-1}, e_h) = C(\bm{x}^{h-2}, e_h)$.
    Applying the same argument until the step $0$ is reached proves the desired statement.
    For the sake of simplifying the indexing, we only consider the second step $h = 2$; the same argument is easily generalizable for the generic step $h \in \{ 2, \dots, p \}$.
    
    We start by unfolding the definition of $C(\bm{x}^1, e_2)$, considering how the walks $\bm{w}^1$ on $G_{\bm{x}^1}$ are originated from the walks $\bm{w}^0$ on $G_{\bm{x}^0}$:
    {\small
    $$
    \everymath{\displaystyle}
    \begin{array}{rl}
        C(\bm{x}^1, e_2) =
             & \sum_{\substack{\bm{w}^1 \\ \text{ walk on } G_{\bm{x}^1}}}
                \sum_{m = 0, 1, 2}
                \theta_m \langle \ \bm{w}^1 \in \cW^{e_2}_m \rangle
                \ \mu^1_{\bm{w}^1} \\
        =
             & \sum_{\bm{w}^0 \in \cW^{e_1}_0} \bigg( \sum_{k=0}^{d_1} \Big( \,
                \sum_{m = 0, 1, 2}
                \theta_m \ \langle \ee^{e_1}_0(\bm{w}^0)_k \in \cW^{e_2}_m \rangle
                \ \mu^1_{\ee^{e_1}_0(\bm{w}^0)_k} \Big) \bigg) \\
             & + \ \sum_{\bm{w}^0 \in \cW^{e_1}_1} \bigg(
                \sum_{m = 0, 1, 2}
                \theta_m \ \langle \ee^{e_1}_1(\bm{w}^0) \in \cW^{e_2}_m \rangle
                \ \mu^1_{\ee^{e_1}_1(\bm{w}^0)} \bigg) \\
             & + \ \sum_{\bm{w}^0 \in \cW^{e_1}_2} \bigg(
                \sum_{m = 0, 1, 2}
                \theta_m \ \langle \ee^{e_1}_2(\bm{w}^0) \in \cW^{e_2}_m \rangle
                \ \mu^1_{\ee^{e_1}_2(\bm{w}^0)} \bigg) \ .
    \end{array}
    $$}
    
    Notice that, when extending a walk on the $1$-edge $e_1$, its multiplicity on $e_2$ is not affected.
    Therefore, for all $m = 0, 1, 2$, we have the equivalent conditions:
    $$
    \begin{array}{ccccccc}
        \bm{w}^0 \in \cW^{e_2}_m                      & \Leftrightarrow
            & \ee^{e_1}_0(\bm{w}^0)_k \in \cW^{e_2}_m & \Leftrightarrow
            & \ee^{e_1}_1(\bm{w}^0) \in \cW^{e_2}_m   & \Leftrightarrow
            & \ee^{e_1}_2(\bm{w}^0) \in \cW^{e_2}_m \ , \\[5pt]
        \langle \bm{w}^0 \in \cW^{e_2}_m \rangle                      & =
            & \langle \ee^{e_1}_0(\bm{w}^0)_k \in \cW^{e_2}_m \rangle & =
            & \langle \ee^{e_1}_1(\bm{w}^0) \in \cW^{e_2}_m \rangle   & =
            & \langle \ee^{e_1}_2(\bm{w}^0) \in \cW^{e_2}_m \rangle \ .
    \end{array}
    $$

    With this observation, the constant $C(\bm{x}^1, e_2)$ may finally be rewritten as $C(\bm{x}^0, e_2)$, thus completing the proof:
    {\small
    $$
    \everymath{\displaystyle}
    \begin{array}{rl}
        C(\bm{x}^1, e_2) =
             & \sum_{\bm{w}^0 \in \cW^{e_1}_0} \bigg( \sum_{k=0}^{d_1} \Big( \,
                \sum_{m = 0, 1, 2}
                \theta_m \ \langle \bm{w}^0 \in \cW^{e_2}_m \rangle
                \ \frac{1}{d_1 \!+\! 1} \mu^0_{\bm{w}^0} \Big) \bigg) \\
             & + \ \sum_{\bm{w}^0 \in \cW^{e_1}_1} \bigg(
                \sum_{m = 0, 1, 2}
                \theta_m \ \langle \bm{w}^0 \in \cW^{e_2}_m \rangle
                \ \mu^0_{\bm{w}^0} \bigg) \\
             & + \ \sum_{\bm{w}^0 \in \cW^{e_1}_2} \bigg(
                \sum_{m = 0, 1, 2}
                \theta_m \ \langle \bm{w}^0 \in \cW^{e_2}_m \rangle
                \ \mu^0_{\bm{w}^0} \bigg) \\
        =
             & \sum_{\substack{\bm{w}^0 \\ \text{ walk on } G_{\bm{x}^0}}}
                \sum_{m = 0, 1, 2}
                \theta_m \langle \ \bm{w}^0 \in \cW^{e_2}_m \rangle
                \ \mu^0_{\bm{w}^0} \\
        =
             & C(\bm{x}^0, e_2) \ .
    \end{array}
    $$}
    \qed
\end{proof}

Finally, at the very end of the procedure, the last dual variables $\bm{\mu}^p$ are adjusted by the global factor
$$
C^*(\bm{x}^0)
:= \max_{h = 1, \dots, p} \{ C(\bm{x}^0, e_h) \}
 = \max_{h = 1, \dots, p} \{ C(\bm{x}^{h-1}, e_h) \} \ .
$$
The claims of this section guarantee that the rescaled assignment \
$\bm{\mu}^* := \frac{1}{C^*(\bm{x}^0)} \bm{\mu}^p$ \
is feasible and attains the objective value \
$\frac{1}{C^*(\bm{x}^0)} \dual\optii(\bm{x}^0)$, \
therefore yielding, by \cref{eq:optp_and_gapp}, \cref{eq:chain_ineq}, and  \Cref{lem:doptp_ii}, the inequality:
$$
\gapp(\bm{x}') \leq C^*(\bm{x}^0) \cdot \gapp(\bm{x}^0) \ .
$$

This final inequality is the one used in the proof of \Cref{thm:GB_algo}, which is now complete in all its details.

%%%%%%%%%%%%%%%%%%%%
% Implementation details for the GB algorithm
%%%%%%%%%%%%%%%%%%%%
\section{Implementation details for the \texorpdfstring{\gb}{GB} algorithm} \label{sec:impl_details}

This section enriches the content of \Cref{sec:comp_results}, discussing implementation details of the \gb\ algorithm.

%%%%%%%%%% Details on the computation of DOPTII
\subsection{Details on the computation of \texorpdfstring{$\dual\optii$}{DOPTII}}

First, we briefly delve into the procedure of solving $\dual\optii(\bm{x})$ \eqref{eq:doptii_obj} -- \eqref{eq:doptii_var_mu}, used as a subroutine of \Cref{alg:GB_algo}.

It is not computationally efficient to solve the dual problem $\dual\optii(\bm{x})$ with all variables included from the outset. Instead, we adopt a well-known \emph{row generation} approach to the primal: constraints are added incrementally, one at a time, until it can be proven that all remaining constraints are implied by those already included. This technique is analogous to the separation of subtour elimination constraints in the standard approach to solving the TSP (See, e.g., \cite{conforti2014integer}).
The LP to solve becomes $\optii(\bm{x})$, the primal of $\dual\optii(\bm{x})$:
variables (respectively constraints) of the former correspond to constraints (respectively variables) of the latter.
\begin{align}
    \optii(\bm{x}) := \
    \text{minimize \ \ }
        & \sum_{ij \in E_{\bm{x}}} x_{ij} c_{ij}
            \notag \\
    \text{subject to: \ }
        & \sum_{ij\in E_{\bm{x}}} w_{ij} c_{ij} \ \geq 1
            & & \forall \bm{w} \text{ walk on $G_{\bm{x}}$},
            \notag \\
        & c_{ij} \geq 0
            & & \forall ij \in E_{\bm{x}}.
            \notag
\end{align}

The constraints associated with the walks on $G_{\bm{x}}$ are added to the model sequentially, choosing the ``most restrictive'', i.e., the ones corresponding to the walks of minimal cost.
These may be found by solving the following ILP, which is the version of the graph-\tsp\ with costs on the edges.

\begin{align}
    \text{minimize \ \ }
        & \sum_{ij \in E_{\bm{x}}} c_{ij} w_{ij}
            \notag \\
    \text{subject to: \ }
        & \sum_{ij \in \delta(v)} w_{ij} \ = 2 \ d_v
            & & \forall v \in V_n,
            \notag \\
        & \sum_{ij \in \delta(S)} w_{ij} \ \geq 2
            & & \forall S \in \cS,
            \notag \\
        & 0 \leq \ w_{ij} \ \leq 2
            & & \forall {ij} \in E_{\bm{x}},
            \notag \\
        & w_{ij} \ \ \text{integer}
            & & \forall {ij} \in E_{\bm{x}},
            \notag \\
        & d_v \ \ \text{integer}
            & & \forall v \in V_n.
            \notag
\end{align}
Once the solution is proven to be optimal, we formulate $\dual\optii$ using only those variables $\mu_{\bm{w}}$ corresponding to tight constraints of $\optii$.
In virtue of the theorem of complementary slackness (see, e.g., \cite{gale1951}), all remaining dual variables can be safely set to zero.
At this point, the objective values of the primal and dual formulations coincide, and we recover the non-zero $\mu_{\bm{w}}$ values required for the estimates $C(\bm{x}, e)$.

%%%%%%%%%% Details on the obtained upper bounds
\subsection{Details on the obtained upper bounds}

To prove \Cref{thm:Gap_4over3}, we executed the \gb\ algorithm on all the ancestors of $\cA_k$ with $k = 3, 4, 5, 6$.
As anticipated in \Cref{sec:comp_results}, these initial runs have not always returned the desired $\nicefrac{4}{3}$ bound.
Consequently, to accomplish our objective, the \gb\ algorithm was also applied to certain successors of these ``unsatisfying'' ancestors, in accordance with \Cref{lem:bound_ancestors}.
Since there is some flexibility in choosing the next successor to explore, we adopted the following heuristic:
given a vertex $\bm{x}$ with $\gb(\bm{x}) > \nicefrac{4}{3}$, we considered the successor $\bm{x}' = \bb(\bm{x}, e)$ obtained by the application of a \bb-move on the $1$-edge $e$ of $\bm{x}$ with the largest constants $C(\bm{x}, e)$ (as computed in \Cref{alg:GB_algo}, line 4).

Notice that, when repeatedly applying the \gb\ algorithm on subsequent successors, the procedure of solving $\dual\optii(\bm{x}')$ (subroutine of \Cref{alg:GB_algo}, line 7) may benefit from the previous solution of $\dual\optii(\bm{x})$, already at our disposal: instead of restarting the row generation technique from zero, we may take the optimal walks of $\bm{x}$, transform them into walks on $\bm{x}'$ as described at the beginning of \Cref{subsec:bounding_gap}, and initialize the model with the associated constraints.

\Cref{tab:comp_res} reports, for every $k = 3, 4, 5, 6$,
the number of ancestors in $\cA_k$ (up to isomorphism),
the upper bound found on the \gap\ for the vertices in $\cF_k$ (namely, the maximum of all the values returned by the \gb\ algorithm applied on the ancestors in $\cA_k$ and the selected successors),
and the maximum number of additional consecutive iterations of the \gb\ algorithm needed on the ancestor to achieve the final bound.

\begin{table}[htbp]
    \centering
    \caption{Computational results of the application of the \gap-Bounding algorithm.}
    \label{tab:comp_res}
    \renewcommand{\arraystretch}{1.5}
    \begin{tabular}{c@{\hskip 15pt}c@{\hskip 15pt}c@{\hskip 15pt}c}
        $k$
            & $|\cA_k|$
            & \begin{tabular}{c}
                upper bound on \gap \\[-5pt]
                for $\cF_k$
            \end{tabular}
            & \begin{tabular}{c}
                max additional iterations \\[-5pt]
                of \gb\ algorithm
            \end{tabular} \\
        \hline
        $3$
            & $1$
            & $\nicefrac{4}{3}$
            & $0$ \\
        $4$
            & $5$
            & $\nicefrac{4}{3}$
            & $2$ \\
        $5$
            & $44$
            & $\nicefrac{4}{3}$
            & $5$ \\
        $6$
            & $715$
            & $\nicefrac{4}{3}$
            & $10$ \\
        \hline
    \end{tabular}
\end{table}

We remark that the values given in the table are upper bounds: in the perspective of proving the conjecture, we were satisfied with $\nicefrac{4}{3}$, but in principle the actual \gap\ on a certain family may be even lower.

\end{document}